\documentclass[a4,11pt]{article}
\usepackage{amsfonts,latexsym,eucal,color,graphicx,epsfig,amssymb,amsmath,bm}
\newcommand{\be}{\begin{eqnarray}}
\newcommand{\ee}{\end{eqnarray}}
\newcommand{\pd}{\partial}

\newcommand{\nn}{\nonumber}
\begin{document}

\begin{center}
{\LARGE {\bf Explosive particle creation by instantaneous \\ change of boundary condition}}\\
\vspace{5mm}
{\bf Umpei Miyamoto}\\
RECCS, Akita Prefectural University, Akita 015-0055, Japan\\
umpei@akita-pu.ac.jp\\
\end{center}

\begin{abstract}
We investigate the dynamic Casimir effect (DCE) of a $1+1$ dimensional free massless scalar field in a finite or semi-infinite cavity for which the boundary condition (BC) instantaneously changes from the Neumann to Dirichlet BC or reversely. While this setup is motivated by the gravitational phenomena such as the formation of strong naked singularities or wormholes, and the topology change of spacetimes or strings in quantum gravity, the analysis is quite general. For the Neumann-to-Dirichlet cases, we find two components of diverging flux emanate from the point where the BC changes. We carefully compare this result with that of Ishibashi and Hosoya (2002) obtained in the context of a quantum version of cosmic censorship hypothesis, and show that one of the diverging components was overlooked by them and is actually non-renormalizable, suggesting to bring non-negligible backreaction or semiclassical instability. On the other hand, for the Dirichlet-to-Neumann cases, we reveal for the first time that only one component of diverging flux emanates, which is the same kind as that overlooked in the Neumann-to-Dirichlet cases. This result suggests not only the robustness of the appearance of diverging flux in instantaneous limits of DCE but also that the type of divergence sensitively depends on the combination of initial and final BCs. 
\end{abstract}

\tableofcontents

\section{Introduction}
\label{sec:intro}

One of the surprising pictures that quantum field theories provide is that a classical vacuum is fluctuating, in which virtual particles spontaneously appear and disappear in short period of time allowed by the Heisenberg uncertainty  principle. The phenomenon that neutral conductive plates put parallel in a classical electromagnetic vacuum attract each other is caused by such particles and called the Casimir effect~\cite{Casimir} (see \cite{lecture} for a review). If one moves the plates (or boundaries for a quantum field in general), the virtual particles can convert into real ones, which is known as the dynamic Casimir effect (DCE)~\cite{Moore} (see \cite{Dalvit:2010ria} for a review).

In general, in order to realize and detect the DCE experimentally by moving a boundary, its speed has to be accelerated up to a few percent of the speed of light. Therefore, such an experiment had been thought to be quite difficult. The effect brought by moving a boundary, however, was recognized to be realized effectively by modulating with high frequency the electromagnetic properties of a static boundary. Based on this idea, the DCE was indeed observed first by Wilson et al.~\cite{nature} using a superconducting circuit. So far, there have been proposed various experimental methods to realize the DCE~\cite{CurrentStatus, Nation:2011dka,Farina:2012qd}, and various theoretical results have been obtained~\cite{Dodonov:2001yb}.

It is mentioned that the dynamics of quantum systems undergoing a rapid change of parameter in their Hamiltonian (note that the change of boundary conditions can be included as terms in the Hamiltonian) is called the quantum quench dynamics and is actively studied nowadays since it poses many fundamental questions that can be studied by current-generation experiments~\cite{Mitra}. For example, the effect of time-periodic boundary condition (i.e., Floquet dynamics) in a conformal field theory, which is a low-energy description of quantum critical systems, has been investigated~\cite{Berdanier:2017kmd}. The entanglement entropy of a conformal field excited by the change of boundary condition (BC) also has been studied~\cite{He:2014mwa}.

While the DEC is a universal phenomenon caused by time-dependent BCs, it occupies a special position in general relativity and other gravitational theories since the effects similar or equivalent to time-dependent BCs are realized not artificially but naturally in dynamical spacetimes, such as expanding universe~\cite{Parker}, the gravitational collapse of stars to black holes~\cite{Hawking:1974sw}, the creation of naked singularities~\cite{Ford:1978ip, Ishibashi:2002ac} and wormholes~\cite{Braunstein:1996aj}, and the topology change of spacetime (and string worldsheets) in quantum gravity~\cite{Anderson:1986ww, Manogue,Shapere:2012wn}  (see \cite{Birrell:1982ix, Wald:1995yp} for a comprehensive study of quantum field theory in curved spacetimes). 

Among the above phenomena in gravitational physics, the particle creation due to the formation of naked singularities~\cite{Ishibashi:2002ac} is of fundamental importance since it is closely related to the future predictability of law of quantum physics, namely, the existence of cosmic censor~\cite{Penrose:1969pc} from the quantum physics point of view. The basic idea is as follow. The spacetime singularity, in which the predictability of law of physics is thought to be lost if the singularity is naked or visible, is defined by the geodesic incompleteness~\cite{Hawking:1973uf}. However, such a definition of singular spacetime using the notion of particles may judge a spacetime appearing harmless (e.g., Minkowski spacetime from which a single point is taken out) to be singular. Therefore, it was proposed to define the spacetime regularity with not the goedesic completeness but a uniqueness of propagation of classical wave fields (or a uniqueness of self-adjoint extension of time-translation operator)~\cite{Wald:1980jn, Horowitz:1995gi}. Such a definition using the notion of fields actually excludes the spacetimes appearing harmless from a class of singular spacetimes~\cite{Ishibashi:1999vw}. 

Ishibashi and Hosoya~\cite{Ishibashi:2002ac} proceeded to a next step. Namely, they investigated what happens if one quantizes a wave field in the `wave-singular' (therefore singular also in the ordinary geodesic sense) spacetime describing the formation of a strong naked singularity, which can be modeled by instantaneous change of BCs for the wave field. More specifically, they considered a quantized $1+1$ dimensional free massless scalar field in a cavity for which the BC suddenly changes from the Neumann to Dirichlet. They showed that a diverging flux taking form of delta function squared emanates from the points where the BCs change and propagates along null lines. From such a result, they concluded that the backreaction of created particles would bring null singularities, resulting in the recovery of global hyperbolicity (i.e., the future predictability of law of physics). That is, the created particles play the role of a quantum version of cosmic censor.

While the idea of the quantum version of cosmic censor is interesting and shown to work in \cite{Ishibashi:2002ac}, an unsatisfactory point may be that the analysis was restricted to the Neumann-to-Dirichlet case. Although Ishibashi and Hosoya tried to examine a more general case for which the BC changes from a Robin BC $\phi(t,x)=a \pd_x \phi$ to another Robin one $\phi(t,x)=b \pd_x \phi$ ($a$ and $b \; (\neq a)$ are constants and the both sides of equalities are evaluated at the boundary), but failed to obtain any rigorous result. (See \cite{Romeo:2000wt} for a systematic study on the static Casimir effect under Robin BCs and \cite{Mintz:2006jh, Mintz:2006yz, Farina:2012qd} for the DCE with time-dependent Robin BCs with a non-relativistic approximation.) 

Therefore, in this paper, we shall extend the analysis in Ref.~\cite{Ishibashi:2002ac} in two directions. First, we examine the instantaneous change of BC in a finite cavity from the Dirichlet to Neumann. Then, we examine both the Neumann-to-Dirichlet (N-D) and Dirichlet-to-Neumann (D-N) cases in a semi-infinite cavity. For the D-N cases both in the finite and semi-infinite cavities, we find with a little surprise that a diverging flux emanates from the point where the BC changes but its property is completely different from that in the N-D case obtained in Ref.~\cite{Ishibashi:2002ac}. Furthermore, in the course that we reproduced the result of N-D case, we found that such a diverging flux appears also in the N-D case in addition to the term of delta function squared, but was overlooked in \cite{Ishibashi:2002ac}. These results suggest that the divergence of flux, which would be a necessary condition for the quantum version of cosmic censor to work, is not a special result in the N-D case. In addition, it is also suggested that the type of divergence sensitively depends on the combination of the initial and final BCs.

Here, let us give a few remarks on the analysis in this paper. The idealization of instantaneous change of BCs, which is natural from the viewpoint of the formation of strong naked singularities, enables us to obtain all the results in analytic form. The particle creation by the rapid appearance and/or disappearance of a wall in a one-dimensional (1D) finite cavity was studied in Refs.~\cite{Rodriguez-Vazquez:2014hka, Brown:2015yma, Harada:2016kkq}.  In particular, the system with the instantaneous appearance and disappearance of a Dirichlet wall studied in \cite{Harada:2016kkq} is more complex than but similar to the system in Sec.~\ref{sec:fin} of the present paper.

The organization of this paper is as follow. In Sec.~\ref{sec:fin}, we investigate the particle creation due to the instantaneous change of BC in a finite 1D cavity, for the N-D case (Sec.~\ref{sec:ND_fin}) and the D-N case (Sec.~\ref{sec:DN_fin}). The origin of discrepancy between the result in Sec.~\ref{sec:fin} and Ref.~\cite{Ishibashi:2002ac} is clarified in Sec.~\ref{sec:ih}. In Sec.~\ref{sec:inf}, the case of semi-infinite cavity is analyzed. We conclude in Sec.~\ref{sec:conc}. The proof of consistency between different quantizations, called the unitarity relations, and some integration formulas are presented in Appendices~\ref{sec:UR} and \ref{sec:int}, respectively. The result for the semi-infinite cavity in Sec.~\ref{sec:inf} is reproduced in Appendix~\ref{sec:green} with the Green-function method, which naturally involves the regularization of the vacuum expectation value of energy-momentum tensor. We work in the natural units in which $c=\hbar=1$.

\section{Finite cavity I}
\label{sec:fin}

\subsection{Quantization of massless scalar field}
\label{sec:quant_fin}

We consider a free massless scalar field in a 1D cavity of which length is $L$,
\be
	(-\pd_t^2+\pd_x^2) \phi(t,x)=0,
\;\;\;
	-\infty < t < \infty,
\;\;\;
	0 < x < L.
\label{eq:eom}
\ee
At the right boundary $x=L$, we assume the homogeneous Dirichlet boundary condition all the time,
\be
	\phi(t,L)=0,
\;\;\;
	-\infty < t < \infty.
\label{eq:bc1}
\ee
At the left boundary $x=0$, we consider two kinds of boundary conditions. One is the homogeneous Neumann boundary condition,
\be
	\pd_x \phi(t,0)=0.
\label{eq:bc2}
\ee
Another is the Dirichlet boundary condition,
\be
	\phi(t,0)
	=
	0.
\label{eq:bc3}
\ee

During boundary conditions \eqref{eq:bc1} and \eqref{eq:bc2} are imposed, a natural set of positive-energy mode functions $\{ f_n \} $ is given by
\be
	f_n (t,x)
	=
	\sqrt{ \frac{2}{n\pi} } e^{-ip_n t} \cos  ( p_n x ) ,
\;\;\;
	p_n := \frac{n \pi}{2L},
\;\;\;
	n = 1,3,5,\cdots.
\label{eq:f_fin}
\ee
In the rest of this paper, we suppose that $n$ and $n'$ entirely denote odd natural numbers, otherwise denoted. The above mode functions satisfy the following orthonormal conditions,
\be
	\langle f_n,f_{n'} \rangle = - \langle f_n^\ast, f_{n'}^\ast \rangle = \delta_{nn'},
\;\;\;
	\langle f_n,f_{n'}^\ast \rangle = 0,
\label{eq:f_ortho_fin}
\ee
where the asterisk denotes the complex conjugate and $\langle \; ,\; \rangle $ denotes the Klein-Gordon inner product~\cite{Birrell:1982ix},
\be
	\langle \phi,\psi \rangle
	:=
	i\int_0^L ( \phi^\ast \pd_t \psi - \pd_t \phi^\ast \psi ) dx.
\label{eq:IP}
\ee

During boundary conditions \eqref{eq:bc1} and \eqref{eq:bc3} are imposed, a natural set of positive-energy mode functions $\{  g_m \} $ is given by
\be
	g_m (t,x)
	=
	\frac{1}{ \sqrt{ m\pi }} e^{-i q_m t} \sin ( q_m x ),
\;\;\;
	q_m := \frac{m \pi}{L},
\;\;\;
	m =1,2,3, \cdots .
\label{eq:g_fin}
\ee
In the rest of this paper, we suppose that $m$ and $m'$ entirely denote natural numbers, otherwise denoted. The above mode functions satisfy the following orthonormal conditions,
\be
	\langle g_m,g_{m'} \rangle = - \langle g_m^\ast, g_{m'}^\ast \rangle = \delta_{mm'},
\;\;\;
	\langle g_m,g_{m'}^\ast \rangle = 0.
\label{eq:g_ortho_fin}
\ee

Associated with the above two sets of mode function, $\{ f_n \}$ and $\{ g_m \}$, there are two ways to quantize the scalar field. One is to expand the scalar field by $f_n$,
\be
	{\bm \phi} = \sum_{\substack{ n=1 \\ n:{\rm odd}}}^\infty
	( {\bm a}_n f_n + {\bm a}_n^\dagger f_n^\ast),
\label{eq:phi_f_fin}
\ee
and impose the commutation relations,
\be
	[ {\bm a}_n, {\bm a}_{n'}^\dagger ] = \delta_{nn'},
\;\;\;
	& [ {\bm a}_n, {\bm a}_{n'} ] = 0.
\label{eq:comm_a_fin}
\ee
By imposing the above commutation relations, the following equal-time canonical commutation relation is realized,
\be
	[ {\bm \phi}(t,x), \pd_t {\bm \phi}(t,x') ] = i \delta (x-x').
\label{eq:canonical}
\ee
Then, ${\bm a}_n$ and ${\bm a}_{n}^\dagger$ are interpreted as the annihilation and creation operators, respectively. The vacuum state in which no particle corresponding to mode function $f_n$ exists is defined by
\be
	{\bm a}_n | 0_f \rangle = 0,
\;\;\;
	n = 1,3,5,\cdots,
\;\;\;
	\langle 0_f | 0_f \rangle = 1.
\label{eq:0f_fin}
\ee

Another is to expand the field by $ g_m$,
\be
	{\bm \phi} = \sum_{m=1}^\infty ( {\bm b}_m g_m + {\bm b}_m^\dagger g_m^\ast),
\label{eq:phi_g_fin}
\ee
and impose the commutation relations,
\be
	[ {\bm b}_m, {\bm b}_{m'}^\dagger ] = \delta_{mm'},
\;\;\;
	 [ {\bm b}_m, {\bm b}_{m'} ] = 0.
\label{eq:comm_b_fin}
\ee
The vacuum state in which no particle corresponding to $g_m$ exists is defined by
\be
	{\bm b}_m | 0_g \rangle = 0,
\;\;\;
	m =1,2,3\cdots,
\;\;\;
	\langle 0_g | 0_g \rangle = 1.
\label{eq:0g_fin}
\ee

Later, we will estimate the vacuum expectation value of energy-momentum tensor for the scalar field. The energy-momentum tensor operator is written as
${\bm T}_{\mu\nu} = \pd_\mu {\bm \phi} \pd_\nu {\bm \phi} -\frac12 \eta_{\mu\nu} ( \pd {\bm \phi} )^2$, where $\eta_{\mu\nu} = {\rm Diag.} (-1,1)$ is the $1+1$ dimensional flat metric. Introducing double null coordinates, non-zero components of this tensor are
\be
	{\bm T}_{\pm\pm} = ( \pd_\pm {\bm \phi} )^2,
\;\;\;
	z_\pm := t \pm x.
\label{eq:em_null}
\ee
Note that the energy density and momentum density in the original Cartesian coordinates are ${\bm T}^{tt}= {\bm T}_{--} + {\bm T}_{++}$ and $ {\bm T}^{tx} = {\bm T}_{--} - {\bm T}_{++}$, respectively.

\subsection{Particle creation by instantaneous change of boundary condition}
\label{sec:creation_fin}

Given the above quantization schemes, we investigate how the vacuum is excited when the boundary condition at left boundary $x=0$ is instantaneously, say at $t=0$, changed from Neumann to Dirichlet (Sec.~\ref{sec:ND_fin}) and reversely (Sec.~\ref{sec:DN_fin}). 

\subsubsection{From Neumann to Dirichlet}
\label{sec:ND_fin}

\begin{figure}
\begin{center}
\begin{minipage}[c]{0.8\textwidth}
\begin{center}
\includegraphics[height=5cm]{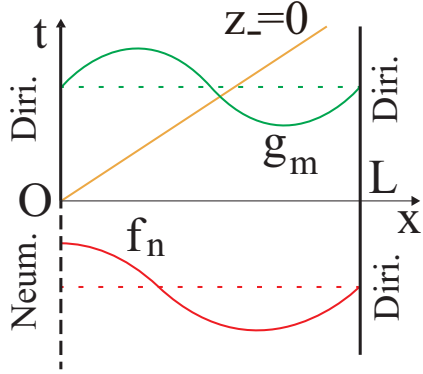}
\caption{The boundary condition at the left end of domain $(x=0)$ instantaneously changes at $t=0$ from Neumann (dashed) to Dirichlet (solid). Spatial configurations of mode functions $f_n$ and $ g_m $ are schematically depicted. }
\label{fig:ND_fin}
\end{center}
\end{minipage}
\end{center}
\end{figure}

First, we assume that the boundary condition at $x=0$ is Neumann~\eqref{eq:bc2} for $t<0$ and Dirichlet~\eqref{eq:bc3} for $t>0$, and that the quantum field is in vacuum $| 0_f \rangle$ in the Heisenberg picture. See Fig.~\ref{fig:ND_fin} for a schematic picture of this situation. Then, we investigate how the vacuum is excited due to the change of boundary condition by computing the spectrum and energy flux of created particles. 

Let us expand $f_n$ by $g_m$,
\be
	f_n
	=
	\sum_{m=1}^\infty ( \alpha_{nm}g_m + \beta_{nm} g_m^\ast ),
\label{eq:fg_fin}
\ee
where the expansion coefficients, called the Bogoliubov coefficients, are computed by
\be
	\alpha_{nm}
	=
	\langle g_m , f_n \rangle,
\;\;\;
	\beta_{nm}
	=
	- \langle g_m^\ast, f_n \rangle.
\label{eq:alpha_beta_form_fin}
\ee
Using the explicit form of mode functions \eqref{eq:f_fin} and \eqref{eq:g_fin}, we obtain
\be
	\alpha_{nm}
	=
	\frac{ 2 }{ (2m-n) \pi } \sqrt{ \frac{ 2m }{ n } },
\;\;\;
	\beta_{nm}
	=
	\frac{ 2 }{ (2m+n) \pi } \sqrt{ \frac{ 2m }{ n } }.
\label{eq:alpha_beta_value_fin}
\ee

Substituting Eq.~\eqref{eq:fg_fin} into Eq.~\eqref{eq:phi_f_fin}, and comparing it with Eq.~\eqref{eq:phi_g_fin}, we obtain
\be
	{\bm b}_m
	=
	\sum_{\substack{ n=1 \\ n:{\rm odd}}}^\infty
	( \alpha_{nm} {\bm a}_n + \beta_{nm}^\ast {\bm a}_n^\dagger ).
\label{eq:ba_fin}
\ee
Substituting Eq.~\eqref{eq:ba_fin} into Eq.~\eqref{eq:comm_b_fin} and using Eq.~\eqref{eq:comm_a_fin}, we obtain
\begin{align}
	\sum_{\substack{ n=1 \\ n:{\rm odd}}}^\infty
	( \alpha_{nm} \alpha_{nm'}^\ast - \beta_{nm}^\ast \beta_{nm'} ) 
	=
	\delta_{mm'},
\;\;\;
	\sum_{\substack{ n=1 \\ n:{\rm odd}}}^\infty
	( \alpha_{nm} \beta_{nm'}^\ast - \beta_{nm}^\ast \alpha_{nm'} ) 
	=
	0,
\label{eq:UR_ND_fin}
\end{align}
which should be satisfied for the two quantizations, Eqs.~\eqref{eq:phi_f_fin} and \eqref{eq:phi_g_fin}, to be consistent. In Appendix~\ref{sec:UR_ND_fin}, these consistency conditions, which we call {\it unitarity relations}, are shown to be satisfied by Bogoliubov coefficients~\eqref{eq:alpha_beta_value_fin}.

The spectrum of created particles is given by the vacuum expectation value of number operator ${\bm b}_m^\dagger {\bm b}_m$,
\be
	\langle 0_f | {\bm b}_m^\dagger {\bm b}_m | 0_f \rangle
	=
	\sum_{\substack{ n=1 \\ n:{\rm odd}}}^\infty
	| \beta_{nm} |^2
	=
	\frac{8}{\pi^2}
	\sum_{\substack{ n=1 \\ n:{\rm odd}}}^\infty
	\frac{ m }{ n ( n+2m )^2 }.
\label{eq:bb_fin}
\ee
Note that this is finite but its summation over $m$, the total number of created particles, is divergent. This implies that the Fock-space representation associated with $ {\bm a}_n $ is unitarily inequivalent to that associated with ${\bm b}_m$~\cite{Wald:1995yp}.

The vacuum expectation value of energy-momentum tensor before the change of boundary condition at $t=0$ is computed by substituting Eq.~\eqref{eq:phi_f_fin} into Eq.~\eqref{eq:em_null}, and using Eqs.~\eqref{eq:comm_a_fin}, \eqref{eq:0f_fin}, and \eqref{eq:f_fin} as
\begin{align}
	\langle 0_f | {\bm T}_{\pm\pm} | 0_f \rangle_{t<0}
	=
	\sum_{\substack{ n=1 \\ n:{\rm odd}}}^\infty
	| \pd_\pm f_n |^2
	=
	\frac{\pi}{8L^2}
	\sum_{\substack{ n=1 \\ n:{\rm odd}}}^\infty n.
\label{eq:ND_t<0_fin}
\end{align}
This represents the Casimir energy density~\cite{Casimir}, which can be made finite with standard regularization schemes~\cite{Birrell:1982ix}.

The most interesting quantity is the vacuum expectation value of energy-momentum tensor after $t=0$. Substituting Eq.~\eqref{eq:phi_g_fin} into Eq.~\eqref{eq:em_null} and using Eq.~\eqref{eq:ba_fin}, we obtain
\begin{gather}
	\langle 0_f | {\bm T}_{\pm\pm} | 0_f \rangle_{t>0}
	=
	\sum_{\substack{ n=1 \\ n:{\rm odd}}}^\infty 
	\sum_{m=1}^\infty \sum_{m'=1}^\infty
	[
		( \alpha_{nm} \beta_{nm'} + \alpha_{nm'} \beta_{nm} )
		{\rm Re}  ( \pd_\pm g_m \pd_\pm g_{m'} )
\nn
\\
		+
		( \alpha_{nm} \alpha_{nm'} + \beta_{nm} \beta_{nm'} )
		{\rm Re} ( \pd_\pm g_m \pd_\pm g_{m'}^\ast )
	].
\label{eq:ND_t>0_form_fin}
\end{gather}
To derive Eq.~\eqref{eq:ND_t>0_form_fin}, we symmetrize it with respect to dummy indices $m$ and $m'$, and use the fact that $\alpha_{nm}$ and $\beta_{nm}$ are real. Using the explicit expressions of Bogoliubov coefficients \eqref{eq:alpha_beta_value_fin} and mode function \eqref{eq:g_fin}, we obtain
\begin{gather}
	\langle 0_f | {\bm T}_{\pm\pm} | 0_f \rangle_{t>0}
=
	\frac{1}{2\pi L^2}
	\sum_{\substack{ n=1 \\ n:{\rm odd}}}^\infty 
	\Bigg(
		\frac{1}{4n}
		[
			4 \sum_{m=1}^\infty \cos ( q_m z_\pm )
			+
			n^2 \sum_{m=1}^\infty \frac{ \cos ( q_m z_\pm ) }{ m^2-(n/2)^2 }
		]^2
		+
		n
		[
			\sum_{m=1}^\infty \frac{ m \sin ( q_m z_\pm ) }{ m^2-(n/2)^2 }
		]^2
	\Bigg).
\label{eq:ND_t>0_value1_fin}
\end{gather}
This is an even function of $z_\pm$ with period $2L$ since it is invariant under reflection $z_\pm \to - z_\pm$ and translation $z_\pm \to z_\pm + 2L$. Therefore, it is sufficient to calculate it in $0 \leq z_\pm < 2L$, and then generalize the obtained expression appropriately to one valid in the entire domain.

The first and second summations over $m$ in Eq.~\eqref{eq:ND_t>0_value1_fin} can be computed to give
\begin{align}
	\langle 0_f | {\bm T}_{\pm\pm} | 0_f \rangle_{t>0}
	=
	\frac{1}{2\pi L^2}
	\sum_{\substack{ n=1 \\ n:{\rm odd}}}^\infty 
	\Bigg(
		\frac{1}{4n}
		[
			16L^2 \delta^2( z_\pm ) + n^2 \pi^2 \sin^2 ( p_n z_\pm )
		]
		+
		n
		[
			\sum_{m=1}^\infty \frac{ m \sin ( q_m z_\pm ) }{ m^2-(n/2)^2 }
		]^2
	\Bigg),
\label{eq:ND_t>0_value2_fin}
\end{align}
which is valid in $  0 \leq z_\pm <2L $, using the following formulas,
\begin{align}
	\sum_{k=1}^\infty \cos( \frac{ 2k \pi}{a} y )
	&=
	-\frac12 + \frac{a}{2} \sum_{\ell = -\infty}^\infty \delta( y- \ell a ),
\;\;\;
	( - \infty  < y < \infty ),
\label{eq:sumForm1}
\\
	\sum_{k=1}^\infty
		\frac{ \cos ky }{ k^2-a^2 }
		&=
		-\frac{\pi}{2a} \cos[ a(\pi - y) ] {\rm cosec} (a \pi )+\frac{1}{2a^2},
\;\;\;
	(0 \leq y \leq 2\pi).
\label{eq:sumForm2}
\end{align}
See Ref.~\cite[p.~730]{maru} for the second formula.

For $z_\pm=0$, from Eq.~\eqref{eq:ND_t>0_value2_fin}, we have
\be
	\langle 0_f | {\bm T}_{\pm\pm} | 0_f \rangle_{t>0}
	=
	\frac{2}{\pi} \sum_{\substack{ n=1 \\ n:{\rm odd}}}^\infty \frac{ \delta^2(0) }{ n },
\;\;\;
	(z_\pm = 0).
\label{eq:ND_t>0_value3_fin}
\ee
For $ 0 <z_\pm < 2L $, the rest summation over $m$ in Eq.~\eqref{eq:ND_t>0_value2_fin} can be computed to give
\be
	\langle 0_f | {\bm T}_{\pm\pm} | 0_f \rangle_{t>0}
	=
	\frac{ \pi }{ 8L^2 } \sum_{\substack{ n=1 \\ n:{\rm odd}}}^\infty n,
\;\;\;
	( 0<z_\pm < 2L ),
\label{eq:ND_t>0_value4_fin}
\ee
using the following formula~\cite[p.~730]{maru},
\begin{align}
	\sum_{k=1}^\infty
	\frac{ k \sin ky }{ k^2-a^2 }
	&=
	\frac{\pi}{2} \sin[ a(\pi - y) ] {\rm cosec} (a \pi ),
\;\;\;
	(0 <y < 2\pi).
\label{eq:sumForm3}
\end{align}

Combining Eqs.~\eqref{eq:ND_t>0_value3_fin} and \eqref{eq:ND_t>0_value4_fin}, we obtain
\be
	\langle 0_f | {\bm T}_{\pm\pm} | 0_f \rangle_{t>0}
	=
	\frac{2}{\pi} \sum_{\substack{ n=1 \\ n:{\rm odd}}}^\infty 
	\frac{ \delta^2( z_\pm ) }{ n }
	+
	\begin{cases}
	0 & (z_\pm =0) \\
	\displaystyle \frac{ \pi }{ 8L^2 } \sum_{\substack{ n=1 \\ n:{\rm odd}}}^\infty  n &  ( 0 < z_\pm < 2L )\\
	\end{cases}.
\label{eq:ND_t>0_value5_fin}
\ee
This is the expression for $ 0 \leq z_\pm < 2L $, what we wanted to know. Extending the domain of Eq.~\eqref{eq:ND_t>0_value5_fin}, we obtain 
\begin{align}
	\langle 0_f | {\bm T}_{\pm\pm} | 0_f \rangle_{t>0}
	=
	\frac{2}{\pi} \sum_{\substack{ n=1 \\ n:{\rm odd}}}^\infty \frac{1}{n}
	\sum_{\ell = -\infty}^\infty \delta^2(z_\pm - 2 \ell L)
	+
	\begin{cases}
	0 & (z_\pm = 2\ell L, \; \ell \in {\bm Z}) \\
	\displaystyle \frac{ \pi }{ 8L^2 } \sum_{\substack{ n=1 \\ n:{\rm odd}}}^\infty n & (\mbox{otherwise}) \\
	\end{cases}.
\label{eq:ND_t>0_value6_fin}
\end{align}

\begin{figure}
\begin{center}
\begin{minipage}[c]{0.8\textwidth}
\begin{center}
	\includegraphics[height=5cm]{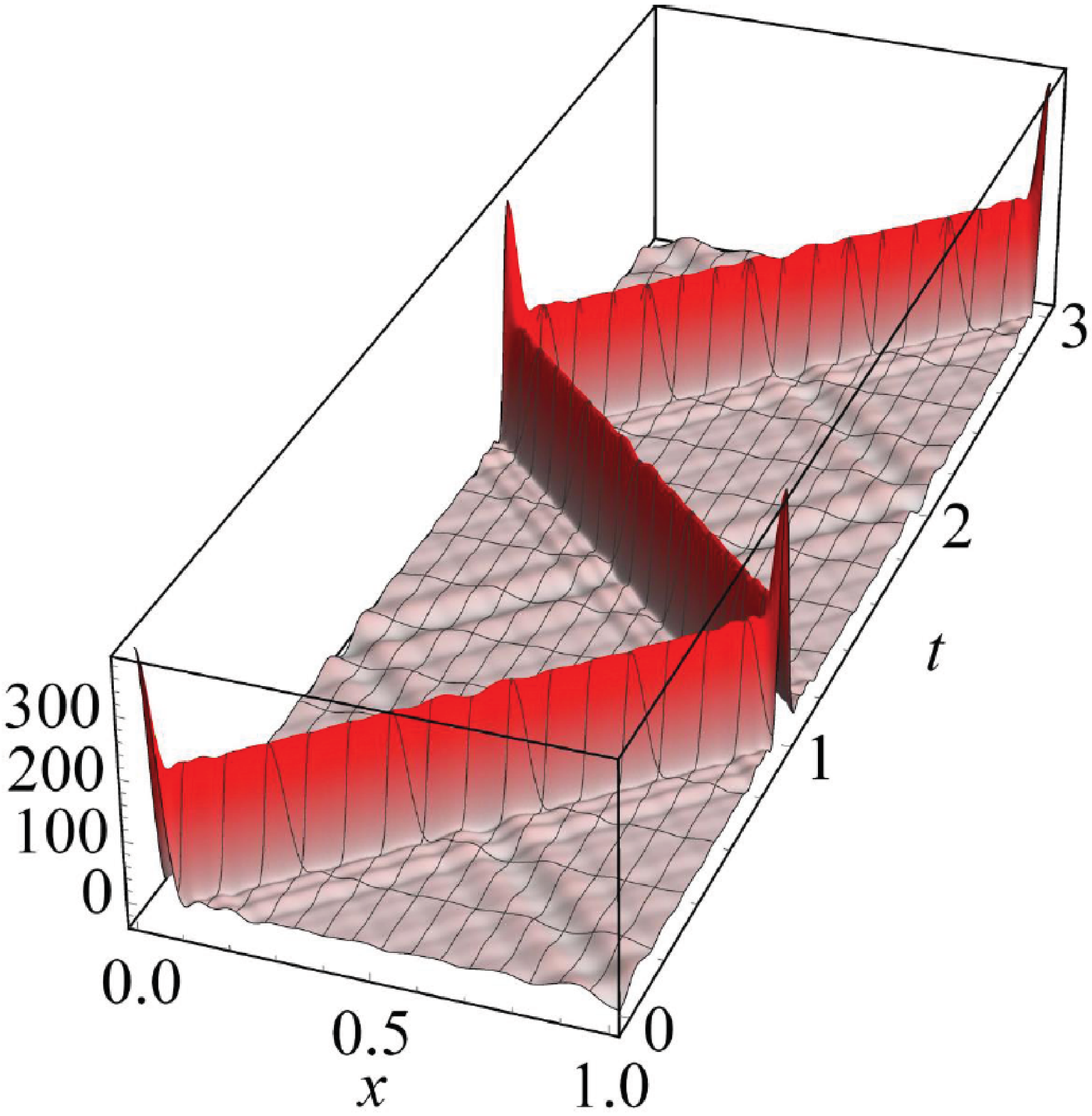} 
	\includegraphics[height=5cm]{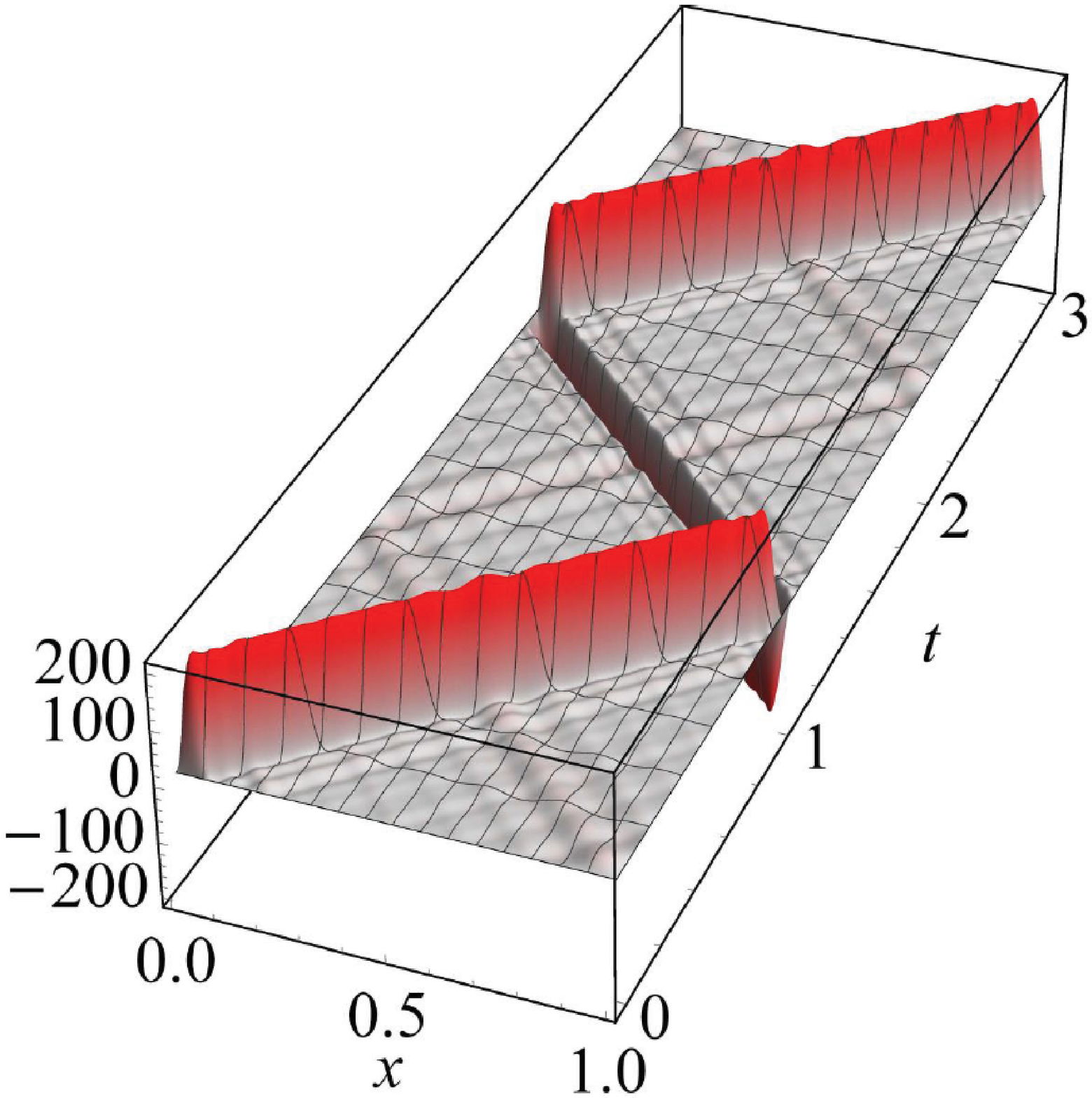}
\caption{Vacuum expectation values of energy density $\langle 0_f | ( {\bm T}_{--} + {\bm T}_{++} ) | 0_f \rangle_{t>0}$ (left) and momentum density $ \langle 0_f | ( {\bm T}_{--} -  {\bm T}_{++}) | 0_f \rangle_{t>0} $ (right) with cutoff, from which the uniform Casimir contribution is subtracted. We set $L=1$ and summation over modes in Eq.~\eqref{eq:ND_t>0_value1_fin} is taken up to $n = m = 13$. The exact results without cutoff are given by Eq.~\eqref{eq:ND_t>0_value6_fin}.}
\label{fig:VEV}
\end{center}
\end{minipage}
\end{center}
\end{figure}

Let us consider the meaning of two terms in  Eq.~\eqref{eq:ND_t>0_value6_fin}. The first term, the delta function squared multiplied by the logarithmically divergent series, represents the diverging flux emanating from the origin $(t,x)=(0,0)$ and localizing on the null lines (Fig.~\ref{fig:VEV}). The dependence of energy density on the delta function squared implies also the divergence of total energy emitted. This component of flux is similar to that predicted in the topology change of 1D universe \cite{Anderson:1986ww} and the same as that predicted in the formation of a strong naked singularity~\cite{Ishibashi:2002ac}.

The second term, at first glance, seems to represent the ambient Casimir energy just like Eq.~\eqref{eq:ND_t<0_fin}, which is negative and finite after a regularization, and its vanishing on the null lines. As will be explicitly shown in the semi-infinite cavity case (see Sec.~\ref{sec:inf} and Appendix~\ref{sec:green}), however, this is not the case. The second term represents the {\it divergence on the null lines} after an appropriate regularization in fact. A simple understanding of such an appearance of divergence is possible as follows. A regularization corresponds to the subtraction of the spatially uniform diverging energy density due to the zero-point oscillation. Therefore, if one subtracts such a uniform diverging quantity from Eq.~\eqref{eq:ND_t>0_value6_fin}, leading to the regularization of ambient Casimir term, a divergence appears {\it on} the null lines $z_\pm = 2\ell L \; (\ell \in {\bm Z})$. 

As far as the present author knows, the second kind of diverging flux was first found in the particle creation due to the instantaneous appearance of Dirichlet wall in a cavity~\cite{Harada:2016kkq}. It was confirmed in the same paper that such a divergence appears in the instantaneous limit of smooth  formation of a Dirichlet wall in cavity analyzed in~\cite{Brown:2015yma}.

It is suspicious that the second kind of flux component does not appear in the analysis of Ishibashi and Hosoya \cite{Ishibashi:2002ac}, since their system is quite similar to the present one. Thus, we will revisit their analysis in Sec.~\ref{sec:ih} and find that the component was overlooked in \cite{Ishibashi:2002ac}.  

\subsubsection{From Dirichlet to Neumann}
\label{sec:DN_fin}

\begin{figure}
\begin{center}
\begin{minipage}[c]{0.8\textwidth}
\begin{center}
\includegraphics[height=5cm]{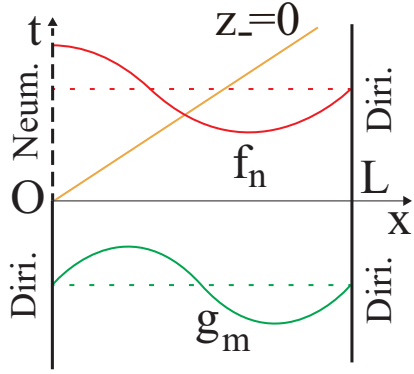}
\caption{The boundary condition at the left end of domain $(x=0)$ instantaneously changes at $t=0$ from Dirichlet (solid) to Neumann (dashed). Spatial configurations of mode functions $ g_m $ and $f_n $  are schematically depicted.}
\label{fig:DN_fin}
\end{center}
\end{minipage}
\end{center}
\end{figure}

We assume that the boundary condition at $x=0$ is Dirichlet~\eqref{eq:bc3} for $t<0$ and Neumann~\eqref{eq:bc2} for $t>0$, and that the quantum field is in vacuum $| 0_g \rangle$. See Fig.~\ref{fig:DN_fin} for a schematic picture of the situation. Since this situation is a kind of time reversal of that in Sec.~\ref{sec:ND_fin}, most parts of calculation can be reused but the results are different.

Let us expand $g_m$ by $f_n$,
\be
	g_m
	=
	\sum_{\substack{ n=1 \\ n:{\rm odd}}}^\infty
	( \rho_{mn} f_n + \sigma_{mn} f_n^\ast ),
\label{eq:gf_fin}
\ee
where the expansion coefficients are given by
\be
	\rho_{mn} =  \langle f_n,g_m \rangle = \alpha_{nm}^\ast,
\;\;\;
	\sigma_{mn} = -\langle f_n^\ast,g_m \rangle = - \beta_{nm}.
\label{eq:rho_sigma_form_fin}
\ee
Here, $\alpha_{nm}$ and $\beta_{nm}$ are given by Eq.~\eqref{eq:alpha_beta_value_fin}. 

Substituting Eq.~\eqref{eq:gf_fin} into Eq.~\eqref{eq:phi_g_fin}, and comparing it with Eq.~\eqref{eq:phi_f_fin}, we obtain
\be
	{\bm a}_n
	=
	\sum_{m=1}^\infty ( \rho_{mn} {\bm b}_m + \sigma_{mn}^\ast {\bm b}_m^\dagger ).
\label{eq:ab_fin}
\ee
Substituting Eq.~\eqref{eq:ab_fin} into Eq.~\eqref{eq:comm_a_fin}, and using Eq.~\eqref{eq:comm_b_fin}, we obtain
\begin{align}
	\sum_{m=1}^\infty ( \rho_{mn} \rho_{mn'}^\ast - \sigma_{mn}^\ast \sigma_{mn'} ) 
	=
	\delta_{nn'},
\;\;\;
	\sum_{m=1}^\infty ( \rho_{mn} \sigma_{mn'}^\ast - \sigma_{mn}^\ast \rho_{mn'} ) 
	=
	0,
\label{eq:UR_DN_fin}
\end{align}
which should be satisfied again for the two quantization, Eqs.~\eqref{eq:phi_f_fin} and \eqref{eq:phi_g_fin}, to be consistent. It is shown in Appendix~\ref{sec:UR_DN_fin} that the Bogoliubov coefficients given by Eq.~\eqref{eq:rho_sigma_form_fin} indeed satisfy unitarity relations \eqref{eq:UR_DN_fin}.

The vacuum expectation value of number operator ${\bm a}_n^\dagger {\bm a}_n$, representing the energy spectrum of created particles, is computed as
\be
	\langle 0_g | {\bm a}_n^\dagger {\bm a}_n | 0_g \rangle
	=
	\sum_{m=1}^\infty | \sigma_{mn} |^2
	=
	\frac{8}{ \pi^2}
	\sum_{m=1}^\infty \frac{ m }{ n(n+2m)^2 }.
\label{eq:aa_fin}
\ee
This and its summation over odd $n$, i.e., the total number of created particles, are divergent. This implies that the Fock-space representation associated with $ {\bm b}_m $ is unitarily inequivalent to that associated with ${\bm a}_n$~\cite{Wald:1995yp}.

The vacuum expectation value of energy-momentum tensor before the change of boundary condition at $t=0$ is computed by substituting Eq.~\eqref{eq:phi_g_fin} into Eq.~\eqref{eq:em_null}, and using the explicit expression of mode function \eqref{eq:g_fin},
\begin{align}
	\langle 0_g | {\bm T}_{\pm\pm} | 0_g \rangle_{t<0}
	=
	\sum_{m=1}^\infty | \pd_\pm g_m |^2
	=
	\frac{ \pi }{ 4L^2 } \sum_{m=1}^\infty m.
\label{eq:DN_t<0_fin}
\end{align}
This represents the Casimir energy density, which can be made finite by standard renormalization procedures~\cite{Birrell:1982ix}.

The vacuum expectation value of energy-momentum tensor after $t=0$ is computed by substituting Eq.~\eqref{eq:phi_f_fin} into Eq.~\eqref{eq:em_null}, and using Eq.~\eqref{eq:ab_fin}, as
\begin{gather}
	\langle 0_g | {\bm T}_{\pm\pm} | 0_g \rangle_{t>0}
	=
	\sum_{m=1}^\infty
	\sum_{\substack{ n=1 \\ n:{\rm odd}}}^\infty
	\sum_{\substack{ n'=1 \\ n':{\rm odd}}}^\infty
	[
		( \rho_{mn} \sigma_{mn'} + \rho_{mn'} \sigma_{mn} )
		{\rm Re} ( \pd_\pm f_n \pd_\pm f_{n'} )
\nn
\\
		+
		( \rho_{mn} \rho_{mn'} + \sigma_{mn} \sigma_{mn'} )
		{\rm Re} ( \pd_\pm f_n \pd_\pm f_{n'}^\ast )
	],
\label{eq:DN_t>0_form_fin}
\end{gather}
which we symmetrize with respect to dummy indices $n$ and $n'$, and we use the fact that $\rho_{mn}$ and $\sigma_{mn}$ are real.

Using the explicit form of Bogoliubov coefficients and mode function, Eqs.~\eqref{eq:rho_sigma_form_fin}, \eqref{eq:alpha_beta_value_fin}, and \eqref{eq:f_fin}, we obtain
\begin{align}
	\langle 0_g | {\bm T}_{\pm\pm} | 0_g \rangle_{t>0}
	=
	\frac{ 4 }{ \pi L^2 } \sum_{m=1}^\infty
	\left(
		4m^3
		[
			\sum_{\substack{ n=1 \\ n:{\rm odd}}}^\infty
			\frac{ \cos ( p_n z_\pm ) }{ n^2-(2m)^2 }
		]^2
		+
		m
		[
			\sum_{\substack{ n=1 \\ n:{\rm odd}}}^\infty
			\frac{ n \sin ( p_n z_\pm ) }{ n^2-(2m)^2 }
		]^2
	\right).
\label{eq:DN_t>0_value1_fin}
\end{align}
This is an even function of $z_\pm$ with period $2L$, since it is invariant under reflection $z_\pm \to -z_\pm$ and translation $z_\pm \to z_\pm + 2L$. Therefore, it is sufficient to calculate it in $0 \leq z_\pm < 2L$, and generalize it appropriately to one valid in the entire domain.

The first summation over odd $n$ in Eq.~\eqref{eq:DN_t>0_value1_fin} can be computed to give
\begin{align}
	\langle 0_g | {\bm T}_{\pm\pm} | 0_g \rangle_{t>0}
	=
	\frac{ 4 }{ \pi L^2 } \sum_{m=1}^\infty
	\left(
		\frac{ m \pi^2 }{ 16 } 
		\sin^2 ( q_m z_\pm )
		+
		m
		[
			\sum_{\substack{ n=1 \\ n:{\rm odd}}}^\infty
			\frac{ n \sin ( p_n z_\pm ) }{ n^2-(2m)^2 }
		]^2
	\right),
\label{eq:DN_t>0_value2_fin}
\end{align}
which is valid in $ 0 \leq z_\pm < 2L $. Here, we have used the following formula~\cite[p.\ 733]{maru},
\begin{align}
	\sum_{k=0}^\infty
	\frac{ \cos [(2k+1)y] }{(2k+1)^2 -a^2} 
	=
	\frac{\pi}{4a} \sin[ \frac{a}{2}(\pi - 2y) ] \sec (\frac{a\pi}{2}),
\;\;\;
	( 0 \leq y \leq \pi ).
\label{eq:maru_typo1}
\end{align}
It is noted here that there are typos in Ref.~\cite[p.\ 733]{maru} about formulas~\eqref{eq:maru_typo1} and \eqref{eq:maru_typo2} (see below).

For $z_\pm=0$, from Eq.~\eqref{eq:DN_t>0_value2_fin}, we have
\be
	\langle 0_g | {\bm T}_{\pm\pm} | 0_g \rangle_{t>0}
	=
	0,
\;\;\;
	( z_\pm = 0 ).
\label{eq:DN_t>0_value3_fin}
\ee
For $ 0 < z_\pm < 2L $, the rest summation over odd $n$ in Eq.~\eqref{eq:DN_t>0_value2_fin} can be computed to give
\begin{align}
	\langle 0_g | {\bm T}_{\pm\pm} | 0_g \rangle_{t>0}
	=
	\frac{ \pi }{ 4 L^2 } \sum_{m=1}^\infty  m
\;\;\;
	( 0 < z_\pm < 2L ),
\label{eq:DN_t>0_value4_fin}
\end{align}
using the following formula~\cite[p.\ 733]{maru},
\begin{align}
	\sum_{k=0}^\infty
	\frac{ (2k+1) \sin [(2k+1)y] }{(2k+1)^2 -a^2} 
	&=
	\frac{\pi}{4} \cos[ \frac{a}{2}(\pi - 2y) ] \sec (\frac{a\pi}{2}),
\;\;\;
	( 0 < y < \pi ).
\label{eq:maru_typo2}
\end{align}

Combining Eqs.~\eqref{eq:DN_t>0_value3_fin} and \eqref{eq:DN_t>0_value4_fin}, and extending the domain periodically into the entire domain, we have
\begin{align}
	\langle 0_g | {\bm T}_{\pm\pm} | 0_g \rangle_{t>0}
	=
	\begin{cases}
		0 & (z_\pm = 2 \ell L, \; \ell \in {\bm Z}) \\
		\displaystyle \frac{ \pi }{ 4L^2 } \sum_{m=1}^\infty m & ({\rm otherwise})
	\end{cases} .
\label{eq:DN_t>0_value5_fin}
\end{align}

Comparing the above result with that in the N-D case~\eqref{eq:ND_t>0_value6_fin}, one sees that there is no flux component of delta function squared in this case. As will be explicitly shown in the semi-infinite cavity case (Sec.~\ref{sec:inf} and Appendix~\ref{sec:green}), Eq.~\eqref{eq:DN_t>0_value5_fin} represents the non-renormalizable diverging flux localized on the null lines $z_\pm = 2 \ell L \; ( \ell \in {\bm Z} )$ and the ambient Casimir energy. Thus, the diverging flux emanates from origin $(t,x)=(0,0)$ and propagates along the null lines in a similar way to Fig.~\ref{fig:VEV}. 

\section{Finite cavity II: Revisit Ishibashi-Hosoya~\cite{Ishibashi:2002ac}}
\label{sec:ih}

As seen in Sec.~\ref{sec:fin}, the vacuum expectation value of energy-momentum tensor has two components in the N-D case as Eq.~\eqref{eq:ND_t>0_value6_fin}, and one component in the D-N case as Eq.~\eqref{eq:DN_t>0_value5_fin}. The origin of such a difference between the N-D and D-N cases will be discussed in Conclusion. Here, let us look into the consistency between these results and a relevant past work.

In Ref.~\cite{Ishibashi:2002ac}, the authors considered the instantaneous change of boundary condition at the both sides of finite cavity. The boundary conditions for $t<0$ are Neumann at the both sides and those for $t>0$ are Dirichlet at the both sides, which we call  the NN-DD case. Since this NN-DD case resembles the N-D case, one can expect the similar results. Namely, we expect that two diverging flux components appear also in the NN-DD case. Reference~\cite{Ishibashi:2002ac}, however, concludes the flux involves only the component of delta function squared. Therefore, we will reconsider here the system adopted in~\cite{Ishibashi:2002ac}, and find that the other component was overlooked. 

\subsection{Quantization of massless scalar field}

\begin{figure}
\begin{center}
\begin{minipage}[c]{0.8\textwidth}
\begin{center}
\includegraphics[height=5cm]{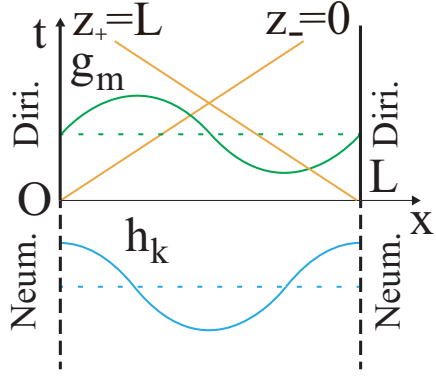}
\caption{The boundary conditions at $x=0$ and $x=L$ are instantaneously changes at $t=0$ from Neumann (dashed) to Dirichlet (solid). Spatial configurations of mode functions $ h_k $ and $g_m$  are schematically depicted.}
\label{fig:NNDD}
\end{center}
\end{minipage}
\end{center}
\end{figure}

We consider the situation that the Neumann boundary condition is imposed at $x=0$ and $x=L$ for $t<0$, while the Dirichlet boundary condition is imposed at $x=0$ and $x=L$ for $t>0$ (see Fig.~\ref{fig:NNDD}). 

In this case, a normalized positive-energy mode function for $t<0$ is given by
\be
	h_k (t,x)
	=
	\frac{1}{\sqrt{k\pi}} e^{-i r_k t} \cos (r_k x),
\;\;\;
	r_k := \frac{k\pi}{L},
\;\;\;
	k =1,2,3,\cdots.
\label{eq:h_ih}
\ee
A normalized mode function for $t>0$ is given by Eq.~\eqref{eq:g_fin}.

The scalar field is quantized by expanding it by set of mode functions $\{ h_k \}$ and an additional zero-mode function ${\bm h}_0$, being spatially uniform, as
\begin{align}
	{\bm \phi}
	=
	{\bm h_0}
	+
	\sum_{k=1}^\infty ( {\bm c}_k h_k + {\bm c}_k^\dagger h_k^\ast ),
\;\;\;
	{\bm h_0}
	=
	\frac{1}{\sqrt{L}}( {\bm Q}+t{\bm P} ).
\label{eq:phi_h_ih}
\end{align}
Here, ${\bm Q}$ and ${\bm P}$ are Hermitian (${\bm Q}^\dagger = {\bm Q}, \; {\bm P}^\dagger = {\bm P}$), and the following commutation relations are imposed  
\be
	[{\bm Q}, {\bm P}] = i,
\;\;\;
	[{\bm Q}, {\bm c}_{k}] = [{\bm P}, {\bm c}_{k}] = 0,
\;\;\;
		[{\bm c}_k, {\bm c}_{k'}^\dagger] = \delta_{kk'},
\;\;\;
	[{\bm c}_k, {\bm c}_{k'}] = 0.
\label{eq:comm_c}
\ee
Note that zero-mode ${\bm h}_0$, which exists because the boundary conditions are Neumann at the both ends, is indispensable to realize the equal-time commutation relation~\eqref{eq:canonical} using commutation relations~\eqref{eq:comm_c}.

\subsection{Particle creation by instantaneous change of boundary condition: From Neumann-Neumann to Dirichlet-Dirichlet}

Let us expand $ {\bm h}_0 $ and  $h_k $ by $g_m$,
\be
	{\bm h}_0 = \sum_{m=1}^\infty ( {\bm \xi}_m g_m + {\bm \xi}_m^\dagger g_m^\ast ),
\;\;\;
	h_k = \sum_{m=1}^\infty ( \xi_{km} g_m + \zeta_{km} g_m^\ast ),
\label{eq:hg}
\ee
where the Bogoliubov coefficients are given by
\be
	{\bm \xi}_m
	=
	\langle g_m, {\bm h}_0 \rangle,
\;\;\;
	\xi_{km}
	=
	\langle g_m, h_k \rangle,
\;\;\;
	\zeta_{km}
	=
	-\langle g_m^\ast, h_k \rangle.
\label{eq:xi_zeta_form_ih}
\ee
Using the explicit form of mode functions~\eqref{eq:g_fin} and \eqref{eq:h_ih}, and Eq.~\eqref{eq:phi_h_ih}, Bogoliubov coefficients~\eqref{eq:xi_zeta_form_ih} are computed as
\begin{gather}
	{\bm \xi}_m
	=
	\frac{ 2 }{ \sqrt{ m\pi L } }
	\left( {\bm Q}+i\frac{L}{m\pi} {\bm P} \right)
	\delta_{m:{\rm odd}},
\label{eq:xi_zeta_value1_ih}
\\
	\xi_{km}
	=
	- \frac{ 2 }{ (k-m)\pi } \sqrt{ \frac{m}{k} } \delta_{k+m:{\rm odd}},
\;\;\;
	\zeta_{km}
	=
	\frac{ 2 }{ (k+m)\pi } \sqrt{ \frac{m}{k} } \delta_{k+m:{\rm odd}}.
\label{eq:xi_zeta_value2_ih}
\end{gather}
Here, we have introduced the following symbols,
\be
	\delta_{k:{\rm odd}}
	:=
	\frac{1-(-1)^k}{2},
\;\;\;
	\delta_{k:{\rm even}}
	:=
	\frac{1+(-1)^k}{2},
\;\;\;
	k \in {\bm Z}.
\label{eq:delta_odd_even}
\ee
Substituting Eq.~\eqref{eq:hg} into Eq.~\eqref{eq:phi_h_ih}, and comparing it with Eq.~\eqref{eq:phi_g_fin}, we have
\be
	{\bm b}_m
	=
	{\bm \xi}_m + \sum_{k=1}^\infty( \xi_{km}{\bm c}_k + \zeta_{km}^\ast {\bm c}_k^\dagger ).
\label{eq:bc_ih}
\ee
Substituting Eq.~\eqref{eq:bc_ih} into Eq.~\eqref{eq:comm_b_fin} and using Eq.~\eqref{eq:comm_c}, we obtain the unitarity relations,
\begin{gather}
\begin{split}
	[ {\bm \xi}_m, {\bm \xi}_{m'}^\dagger ]
	+
	\sum_{k=1}^\infty
	( \xi_{km} \xi_{km'}^\ast - \zeta_{km}^\ast \zeta_{km'} ) = \delta_{mm'},
\\
	[ {\bm \xi}_m, {\bm \xi}_{m'} ]
	+
	\sum_{k=1}^\infty
	( \xi_{km} \zeta_{km'}^\ast - \zeta_{km}^\ast \xi_{km'} ) = 0.
\label{eq:UR_ih}
\end{split}
\end{gather}
In Appendix~\ref{sec:UR_ih}, we will show that the operators given in Eqs.~\eqref{eq:xi_zeta_value1_ih} and \eqref{eq:xi_zeta_value2_ih} satisfy unitarity relations~\eqref{eq:UR_ih}.

We define the vacuum in which no particle corresponding to ${\bm h}_0$ or $h_k$ exist,
\be
	{\bm P} |0_h \rangle = {\bm c}_k |0_h \rangle = 0,
\;\;\;
	k =1,2,3,\cdots.
\ee
Then, the spectrum of created particles are given by the expectation value of number operator $ {\bm b}_m^\dagger {\bm b}_m $,
\begin{align}
&	\langle 0_h | {\bm b}_m^\dagger {\bm b}_m | 0_h \rangle
	=
	\langle 0_h | {\bm \xi}_m^\dagger {\bm \xi}_m | 0_h \rangle
	+
	\sum_{k=1}^\infty | \zeta_{km} |^2
\nn
\\
&	=
	\frac{ 4 }{ m^2\pi^2 }
	\left(
		\frac{m\pi}{L} \langle 0_h | {\bm Q}^2 | 0_h \rangle -1
	\right) \delta_{m:{\rm odd}}
	+
	\frac{ 4 }{ \pi^2 } \sum_{k=1}^\infty \frac{ m }{ k(k+m)^2 } \delta_{k+m:{\rm odd}}.
\label{eq:bb}
\end{align}

The vacuum expectation value of energy-momentum tensor before the change of boundary conditions at $t=0$ is computed  by substituting Eq.~\eqref{eq:phi_h_ih} into Eq.~\eqref{eq:em_null}, and using explicit form of mode function \eqref{eq:h_ih} as
\be
	\langle 0_h | {\bm T}_{\pm \pm} | 0_h \rangle_{t<0}
	=
	\sum_{k=1}^\infty | \pd_\pm h_k |^2
	=
	\frac{\pi}{4L^2} \sum_{k=1}^\infty k.
\ee
This represents the Casimir energy density, which can be made finite by standard regularization schemes such as the $\zeta$-function regularization, the point-splitting regularization, and so on~\cite{Birrell:1982ix}.

The vacuum expectation value of energy-momentum tensor after $t=0$ is computed by substituting Eq.~\eqref{eq:phi_g_fin} into Eq.~\eqref{eq:em_null}, and using Eq.~\eqref{eq:bc_ih},
\begin{gather}
	\langle 0_h | {\bm T}_{\pm \pm} | 0_h \rangle_{t>0}
	=
	\sum_{\substack{ m=1 \\ m:{\rm odd}}}^\infty
	\sum_{\substack{ m'=1 \\ m':{\rm odd}}}^\infty
	\Big[
		\frac{ 8 \langle 0_h | {\bm Q}^2 | 0_h \rangle 
 }{ \pi L \sqrt{ mm' }}
	{\rm Re}
		(
			\pd_\pm g_m \pd_\pm g_{m'} + \pd_\pm g_m \pd_\pm g^\ast_{m'}
		)
\nn
\\
		 +
		\frac{4i}{\sqrt{ \pi^2 m^3m'^{3} }}
		{\rm Im}
			[
				(m+m') \pd_\pm g_m \pd_\pm g_{m'} 
				-
				(m-m') \pd_\pm g_m \pd_\pm g^\ast_{m'} 
			]
	\Big]
\nn
\\
	+
	\sum_{k=1}^\infty \sum_{m=1}^\infty \sum_{m'=1}^\infty
	\Big[
	( \xi_{km} \zeta_{km'} + \zeta_{km}\xi_{km'} ){\rm Re}( \pd_\pm g_m \pd_\pm g_{m'})
	+
	( \xi_{km} \xi_{km'} + \zeta_{km}\zeta_{km'} ){\rm Re}( \pd_\pm g_m \pd_\pm g^\ast_{m'})
	\Big],
\label{eq:NNDD_t>0_form_ih}
\end{gather}
which we symmetrize with respect to dummy indices $m$ and $m'$, and we have used the fact that $\xi_{km}$ and $\zeta_{km}$ are real.

Using explicit form of mode functions~\eqref{eq:g_fin} and Bogoliubov coefficients~\eqref{eq:xi_zeta_value2_ih}, we obtain
\begin{gather}
	\langle 0_h | {\bm T}_{\pm \pm} | 0_h \rangle_{t>0}
	=
	\frac{ 4 \langle 0_h | {\bm Q}^2 | 0_h \rangle }{ L^3 }
	[ \sum_{\substack{ m=1 \\ m:{\rm odd}}}^\infty \cos( q_m z_\pm ) ]^2
	-
	\frac{ 4 i }{ \pi L^2 }
	\sum_{\substack{ m=1 \\ m:{\rm odd}}}^\infty \frac{ \sin( q_m z_\pm ) }{ m }
	\sum_{\substack{ m=1 \\ m:{\rm odd}}}^\infty \cos( q_m z_\pm )
\nn
\\
	+
	\frac{ 4 }{ \pi L^2 } \sum_{\substack{ k=1 \\ k:{\rm odd}}}^\infty	
	\Big(
		\frac{1}{k}
		[
			 \sum_{\substack{ m=2 \\ m:{\rm even}}}^\infty \cos( q_m z_\pm )
			 +
			 k^2
			 \sum_{\substack{ m=2 \\ m:{\rm even}}}^\infty \frac{ \cos( q_m z_\pm )}{m^2-k^2} ]^2
		+
		k [ \sum_{\substack{ m=2 \\ m:{\rm even}}}^\infty \frac{m \sin( q_m z_\pm )}{m^2-k^2} ]^2
	\Big)
\nn
\\
+
	\frac{ 4 }{ \pi L^2 } \sum_{\substack{ k=2 \\ k:{\rm even}}}^\infty	
	\Big(
			\frac{1}{k}
		[
			 \sum_{\substack{ m=1 \\ m:{\rm odd}}}^\infty \cos( q_m z_\pm )
			 +
			 k^2
			 \sum_{\substack{ m=1 \\ m:{\rm odd}}}^\infty \frac{ \cos( q_m z_\pm )}{m^2-k^2} ]^2
		+
		k [ \sum_{\substack{ m=1 \\ m:{\rm odd}}}^\infty \frac{m \sin( q_m z_\pm )}{m^2-k^2} ]^2
	\Big).
\label{eq:NNDD_t>0_value1_ih}
\end{gather}
The summations over odd $m$ in the first two terms of Eq.~\eqref{eq:NNDD_t>0_value1_ih}, both of which are the contributions of the zero-mode, are computed using the following formulas,
\begin{align}
	\sum_{\substack{ k=1 \\ k:{\rm odd}}}^\infty
	\frac{1}{k} \sin ( \frac{ 2k\pi }{ a }y ) 
	=&
	\frac{\pi}{4} \sum_{\ell=-\infty}^\infty (-1)^\ell \Pi_0^{a/2} (y- \frac{a}{2}\ell),
\;\;\;
	( -\infty < y < \infty ),
\label{eq:sumForm1_ih}
\\
	\sum_{\substack{ k=1 \\ k:{\rm odd}}}^\infty
	\cos ( \frac{2k\pi}{a} y )
	=&
	\frac{a}{4} \sum_{\ell=-\infty}^\infty (-1)^\ell \delta (y-\frac{a}{2}\ell ),
\;\;\;
	( -\infty < y < \infty ),
\label{eq:sumForm2_ih}
\end{align}
where $ \Pi_a^b (x) $ is the rectangular function defined as
\be
	\Pi_a^b (x)
	:=
	\int_a^b \delta (x-y)dy
	=
	\begin{cases}
		0 & (x<a, \; b<x) \\
		\frac12 & (x=a, b) \\
		1 & (a<x<b)
	\end{cases}.
\ee
The rest summations over odd and even $m$ in Eq.~\eqref{eq:NNDD_t>0_value1_ih} are computed using formulas~\eqref{eq:sumForm1}, \eqref{eq:sumForm2}, \eqref{eq:sumForm3}, \eqref{eq:maru_typo1}, and \eqref{eq:maru_typo2} in addition to the above formulas, to obtain
\begin{align}
	\langle 0_h | {\bm T}_{\pm \pm} | 0_h \rangle_{t>0}
	=
	\left(
		\frac{ \langle 0_h | {\bm Q}^2 | 0_h \rangle }{ L }
		+
		\frac{1}{\pi} \sum_{k=1}^\infty \frac{1}{k}
	\right)
	\sum_{\ell=-\infty}^\infty \delta^2 (z_\pm - \ell L )
	+
	\begin{cases}
		0 & ( z_\pm  = \ell L,  \ell \in {\bm Z}) \\
		\displaystyle \frac{\pi}{4L^2} \sum_{k=1}^\infty k & ({\rm otherwise})
	\end{cases}.
\label{eq:NNDD_t>0_value2_ih}
\end{align}

After setting $L=\pi$ and regularizing the diverging summation as $\sum_{k=1}^\infty k = -\frac{1}{12}$ by the $\zeta$-function regularization, Eq.~\eqref{eq:NNDD_t>0_value2_ih} should be equal to Eq.~(31) of Ref.~\cite{Ishibashi:2002ac}. The vanishing of Casimir energy on the null lines in Eq.~\eqref{eq:NNDD_t>0_value2_ih}, however, has no counterpart in Eq.~(31) of Ref.~\cite{Ishibashi:2002ac}.
As pointed out at the end of Sec.~\ref{sec:ND_fin}, it should be stressed again that the second term in Eq.~\eqref{eq:NNDD_t>0_value2_ih} represents both the ambient Casimir energy and {\it the divergent flux on the null lines} $( z_\pm  = \ell L,  \ell \in {\bm Z})$ after an appropriate regularization (see Sec.~\ref{sec:inf} and Appendix~\ref{sec:green}), rather than a constant correction to the first divergent term.

While we have derived Eq.~\eqref{eq:NNDD_t>0_value2_ih} with keeping the parallelism with the other analyses in the present paper, it is unclear from where the discrepancy comes. In the next subsection, therefore, we will re-derive Eq.~\eqref{eq:NNDD_t>0_value2_ih} with a method similar to one in Ref.~\cite{Ishibashi:2002ac}.

\subsection{Origin of discrepancy}

Substituting Eq.~\eqref{eq:phi_g_fin} into Eq.~\eqref{eq:em_null}, and using Eq.~\eqref{eq:bc_ih}, the vacuum expectation value of the energy-momentum tensor after $t=0$ is written as
\begin{gather}
	\langle 0_h | {\bm T}_{\pm\pm} | 0_h \rangle_{t>0}
	=
	\sum_{m=1}^\infty \sum_{m'=1}^\infty
	\Big[
	(
		\langle 0_h | {\bm \xi}_m {\bm \xi}_{m'} | 0_h \rangle
		+
		\sum_{k=1}^\infty \xi_{km} \zeta_{km'}^\ast
	) \pd_\pm g_m \pd_\pm g_{m'}
\nn
\\
	+
	(
		\langle 0_h | {\bm \xi}_m {\bm \xi}_{m'}^\dagger | 0_h \rangle
		+
		\sum_{k=1}^\infty \xi_{km} \xi_{km'}^\ast
	) \pd_\pm g_m \pd_\pm g_{m'}^\ast
+
		(
		\langle 0_h | {\bm \xi}_m^\dagger {\bm \xi}_{m'} | 0_h \rangle
		+
		\sum_{k=1}^\infty \zeta_{km} \zeta_{km'}^\ast
	) \pd_\pm g_m^\ast \pd_\pm g_{m'}
\nn
\\
	+
	(
		\langle 0_h | {\bm \xi}_m^\dagger {\bm \xi}_{m'}^\dagger | 0_h \rangle
		+
		\sum_{k=1}^\infty \zeta_{km} \xi_{km'}^\ast
	) \pd_\pm g_m^\ast \pd_\pm g_{m'}^\ast
	\Big].
\label{eq:NNDD_t>0_value3_ih}
\end{gather}
Using explicit form of Bogoliubov coefficients \eqref{eq:xi_zeta_value1_ih} and \eqref{eq:xi_zeta_value2_ih}, and mode function~\eqref{eq:g_fin}, this quantity is rewritten in a compact form,
\begin{gather}
	\langle 0_h | {\bm T}_{\pm\pm} | 0_h \rangle_{t>0}
	=
	\frac{1}{L^3}
	\sum_{\substack{ m=-\infty \\ m:{\rm odd}}}^\infty
	\left( \langle 0_h | {\bm Q}^2 | 0_h \rangle + \frac{ L }{ m\pi } \right) e^{-iq_m z_\pm }
	\sum_{\substack{ m'=-\infty \\ m':{\rm odd}}}^\infty e^{-iq_{m'}z_\pm }
\nn
\\
	+
	\frac{1}{\pi L^2} \sum_{k=1}^\infty \frac{1}{k}
	\sum_{\substack{ m=-\infty \\ m:{\rm odd}}}^\infty
	\frac{ me^{ -i (q_m - q_k ) z_\pm } }{ m-k }  \delta_{m-k:{\rm odd}}
	\sum_{\substack{ m'=-\infty \\ m':{\rm odd}}}^\infty
	\frac{ m'e^{ -i (q_{m'} + q_k ) z_\pm }  }{ m'+k }  \delta_{m'+k:{\rm odd}} .
\label{eq:NNDD_t>0_value4_ih}
\end{gather}
The summations over odd $m$ and $m'$ in Eq.~\eqref{eq:NNDD_t>0_value4_ih} can be evaluated with the following formulas,
\begin{align}
	\sum_{\substack{ k=-\infty \\ k:{\rm odd}}}^\infty
	\frac{1}{k} \exp( -i \frac{ 2k \pi }{ a } y )
	&=
	-\frac{ i \pi }{2} 
	\sum_{\ell=-\infty}^\infty (-1)^\ell \Pi_0^{a/2} (y-\frac{a}{2} \ell),
\label{eq:sumForm3_ih}
\\
	\sum_{\substack{ k=-\infty \\ k:{\rm odd}}}^\infty
	\exp( -i \frac{ 2k \pi }{ a } y )
	&=
	\frac{a}{2}\sum_{\ell=-\infty}^\infty (-1)^\ell \delta (y-\frac{a}{2} \ell),
\label{eq:sumForm4_ih}
\end{align}
which are equivalent to Eqs.~\eqref{eq:sumForm1_ih} and \eqref{eq:sumForm2_ih}, respectively. 

Finally, in order to obtain the final result, it is necessary to use the following relation,
\begin{gather}
	\sum_{\ell=-\infty}^\infty (-1)^\ell \Pi_0^L (z_\pm - \ell L)
	\sum_{\ell'=-\infty}^\infty (-1)^{\ell'} \Pi_0^L (z_\pm - \ell' L)
	=
	\begin{cases}
		0 & (z_\pm = \ell L, \; \ell \in {\bm Z} ) \\
		1 & ({\rm otherwise}) \\
	\end{cases}.
\label{eq:PiSquared}
\end{gather}
Then, we obtain Eq.~\eqref{eq:NNDD_t>0_value2_ih}. It seems that Ref.~\cite{Ishibashi:2002ac} overlooked the fact that the left-hand side of Eq.~\eqref{eq:PiSquared} vanishes on null lines $ z_- = 0$ and $ z_+ = L $. This would be the origin of the discrepancy between our result and their result.

\section{Semi-infinite cavity}
\label{sec:inf}

In the rest of this paper, we investigate the particle creation by the instantaneous change of boundary condition in a semi-infinite cavity, which corresponds to the limit $L \to +\infty$ of the finite-cavity model in Sec.~\ref{sec:fin}. We will see that some simplifications happen in such a limit. Namely, one needs just some simple integral formulas rather than the non-trivial summation formulas in Sec.~\ref{sec:fin}. The analysis in semi-infinite space $x \in [0,+\infty)$ can be a footing to generalize the present analysis, for example, to higher-dimensional models by regarding the spatial coordinate $x$ as a radial coordinate of higher-dimensional spaces (see \cite{Zhou:2016hsh} for a relevant higher-dimensional consideration). While the Bogoliubov transformation will be used in this section again in order to keep the parallelism with the previous sections, the results will be re-derived in Appendix~\ref{sec:green} with an independent method using the Green functions, which naturally involves the point-splitting regularization of the vacuum expectation value of energy-momentum tensor.

\subsection{Quantization of massless scalar field}
\label{sec:quant_inf}

We consider a free massless scalar field in the semi-infinite cavity, of which equation of motion is given by Eq.~\eqref{eq:eom} with $L \to +\infty$.

At left boundary $x=0$, we consider two kinds of boundary conditions. One is the Neumann boundary condition~\eqref{eq:bc2}. Another is the Dirichlet boundary condition~\eqref{eq:bc3}.

During Neumann boundary condition \eqref{eq:bc2} is satisfied, a natural set of positive-energy mode functions $\{ f_p \} $, which is labeled by continuous parameter $p$, is given by
\be
	f_p (t,x)
	=
	\frac{1}{ \sqrt{ \pi p } }e^{ -ipt } \cos  ( p x ) ,
\;\;\;
	p > 0.
\label{eq:f_inf}
\ee
This mode function satisfies the following orthonormal conditions,
\be
	\langle f_p, f_{p'} \rangle = - \langle f_p^\ast, f_{p'}^\ast \rangle = \delta ( p-p' ),
\;\;\;
	\langle f_p, f_{p'}^\ast \rangle = 0,
\label{eq:f_ortho_inf}
\ee
where the integration range of Klein-Gordon inner product, Eq.~\eqref{eq:IP}, is from $0$ to $+\infty$.

During Dirichlet boundary condition \eqref{eq:bc3} is satisfied, a natural set of positive-energy mode functions $ \{ g_q \} $ is given by
\be
	g_q (t,x)
	=
	\frac{1}{ \sqrt{ \pi q } } e^{-i q t} \sin  ( q x ),
\;\;\;
	q > 0.
\label{eq:g_inf}
\ee
This mode function satisfies the following orthonormal conditions,
\be
	\langle g_q, g_{q'} \rangle = - \langle g_q^\ast, g_{q'}^\ast \rangle = \delta (q-q'),
\;\;\;
	\langle g_q,g_{q'}^\ast \rangle = 0.
\label{eq:g_ortho_inf}
\ee

Associated with the above two sets of mode functions, $\{ f_p \}$ and $\{ g_q \}$, there are two ways to quantize the scalar field. Namely, we can expand the scalar field by two sets of mode functions,
\begin{align}
	{\bm \phi}
	&=
	\int_0^\infty dp ( {\bm a}_p f_p + {\bm a}_p^\dagger f_p^\ast),
\label{eq:phi_f_inf}
\\
	{\bm \phi}
	&=
	\int_0^\infty dq ( {\bm b}_q g_q + {\bm b}_q^\dagger g_q^\ast),
\label{eq:phi_g_inf}
\end{align}
where the expansion coefficients are imposed the commutation relations,
\begin{align}
	[ {\bm a}_p, {\bm a}_{p'}^\dagger ] = \delta (p-p'),
\;\;\;
	& [ {\bm a}_p, {\bm a}_{p'} ] = 0,
\label{eq:comm_a_inf}
\\
	[ {\bm b}_q, {\bm b}_{q'}^\dagger ] = \delta (q-q') ,
\;\;\;
	& [ {\bm b}_q, {\bm b}_{q'} ] = 0.
\label{eq:comm_b_inf}
\end{align}
Operators ${\bm a}_p$ and ${\bm b}_q$ (resp.\ ${\bm a}^\dagger_p$ and ${\bm b}^\dagger_q$) are interpreted as annihilation (resp.~creation) operators. 

Accordingly, we can define two normalized vacuum states,
\begin{align}
	{\bm a}_p | 0_f \rangle = 0,
\;\;\;
	& \forall p >0,
\;\;\;
	\langle 0_f | 0_f \rangle = 1,
\label{eq:0f_inf}
\\
	{\bm b}_q | 0_g \rangle = 0,
\;\;\;
	& \forall q >0,
\;\;\;
	\langle 0_g | 0_g \rangle = 1.
\label{eq:0g_inf}
\end{align}
Then, $| 0_f \rangle$ (resp.~$| 0_g \rangle$) is the state where no particle corresponding to $f_n$ (resp.~$g_m$) exists.

\subsection{Particle creation by instantaneous change of boundary condition}
\label{sec:creation_inf}

Given the above quantization of scalar field in the semi-infinite cavity, we investigate how the vacuum is excited when the boundary condition at $x=0$ instantaneously changes from Neumann to Dirichlet (N-D) in Sec.~\ref{sec:ND_inf} and reversely (D-N) in Sec.~\ref{sec:DN_inf}. 

\subsubsection{From Neumann to Dirichlet}
\label{sec:ND_inf}

\begin{figure}
\begin{center}
\begin{minipage}[c]{0.8\textwidth}
\begin{center}
\includegraphics[height=5cm]{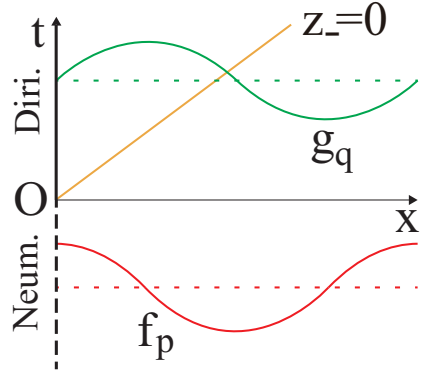}
\caption{The boundary condition at the left end of domain $(x=0)$ instantaneous changes at $t=0$ from Neumann (dashed) to Dirichlet (solid). Spatial configurations of mode functions $ f_p $ and $g_q$  are schematically depicted.}
\label{fig:ND_inf}
\end{center}
\end{minipage}
\end{center}
\end{figure}

We assume that the boundary condition at $x=0$ is Neumann~\eqref{eq:bc2} for $t<0$ and Dirichlet~\eqref{eq:bc3} for $t>0$, and that the quantum field is in vacuum $| 0_f \rangle$, defined by Eq.~\eqref{eq:0f_inf}. See Fig.~\ref{fig:ND_inf} for a schematic picture of the situation.

Let us expand $f_p$ by $g_q$ as,
\be
	f_p
	=
	\int_0^\infty dq ( \alpha_{pq} g_q + \beta_{pq} g_q^\ast ),
\label{eq:fg_inf}
\ee
where the expansion coefficients are given by
\be
	\alpha_{pq}
	=
	\langle g_q , f_p \rangle,
\;\;\;
	\beta_{pq}
	=
	- \langle g_q^\ast, f_p \rangle.
\label{eq:alpha_beta_form_inf}
\ee
Using Eqs.~\eqref{eq:f_inf} and \eqref{eq:g_inf}, we obtain
\be
	\alpha_{pq}
	=
	- \frac{ 1 }{ (p-q) \pi } \sqrt{ \frac{ q }{ p } },
\;\;\;
	\beta_{pq}
	=
	\frac{ 1 }{ (p+q) \pi } \sqrt{ \frac{ q }{ p } },
\label{eq:alpha_beta_value_inf}
\ee
where we have used integral formula $\int_0^\infty e^{iax} dx = ia^{-1} \; ( -\infty < a <\infty)$.

Substituting Eq.~\eqref{eq:fg_inf} into Eq.~\eqref{eq:phi_f_inf}, and comparing it with Eq.~\eqref{eq:phi_g_inf}, we obtain
\be
	{\bm b}_q
	=
	\int_0^\infty dp ( \alpha_{pq} {\bm a}_p + \beta_{pq}^\ast {\bm a}_p^\dagger ).
\label{eq:ba_inf}
\ee
Substituting Eq.~\eqref{eq:ba_inf} into Eq.~\eqref{eq:comm_b_inf}, and using Eq.~\eqref{eq:comm_a_inf}, we obtain the unitarity relations,
\begin{align}
	\int_0^\infty dp ( \alpha_{pq} \alpha_{pq'}^\ast - \beta_{pq}^\ast \beta_{pq'} ) 
	=
	\delta (q-q'),
\;\;\;
	\int_0^\infty dp ( \alpha_{pq} \beta_{pq'}^\ast - \beta_{pq}^\ast \alpha_{pq'} ) 
	=
	0.
\label{eq:UR_ND_inf}
\end{align}
In Appendix~\ref{sec:UR_ND_inf}, we prove that Bogoliubov coefficients \eqref{eq:alpha_beta_value_inf} satisfy Eq.~\eqref{eq:UR_ND_inf}. 

The spectrum of created particles are computed as
\be
	\langle 0_f | {\bm b}_q^\dagger {\bm b}_q | 0_f \rangle
	=
	\int_0^\infty dp | \beta_{pq} |^2
	=
	\frac{1}{\pi^2}
	\int_0^\infty dp \frac{ q }{ p ( p+q )^2 }.
\label{eq:bb_inf}
\ee
This and its integration over $q$ are divergent due to the contribution from the infrared regime.

The vacuum expectation value of energy-momentum tensor before the change of boundary condition at $t=0$ is computed by substituting Eq.~\eqref{eq:phi_f_inf} into Eq.~\eqref{eq:em_null}, and using Eqs.~\eqref{eq:comm_a_inf} and \eqref{eq:f_inf}, as
\begin{align}
	\langle 0_f | {\bm T}_{\pm\pm} | 0_f \rangle_{t<0}
	=
	\int_0^\infty dp | \pd_\pm f_p |^2
	=
	\frac{1}{4\pi} \int_0^\infty dp p.
\label{eq:ND_t<0_inf}
\end{align}
Unlike the finite-cavity case, there is no Casimir energy in this semi-infinite case. The above result just represents the divergent energy density due to the zero-point oscillation. Thus, the renormalized vacuum expectation value obtained by subtracting such a zero-point contribution identically vanishes everywhere as Eq.~\eqref{VEVf_ren2}.

The vacuum expectation value of energy-momentum tensor after $t=0$ is computed by substituting Eq.~\eqref{eq:phi_g_inf} into Eq.~\eqref{eq:em_null}, and using Eq.~\eqref{eq:ba_inf}, as
\begin{gather}
	\langle 0_f | {\bm T}_{\pm\pm} | 0_f \rangle_{t>0}
	=
	\int_0^\infty \int_0^\infty \int_0^\infty dp dq dq'
	[
		( \alpha_{pq} \beta_{pq'} + \alpha_{pq'} \beta_{pq} )
		{\rm Re} ( \pd_\pm g_q \pd_\pm g_{q'} )
\nn
\\
		+
		( \alpha_{pq} \alpha_{pq'} + \beta_{pq} \beta_{pq'} )
		{\rm Re} ( \pd_\pm g_q \pd_\pm g_{q'}^\ast )
	].
\label{eq:ND_t>0_form_inf}
\end{gather}
To derive Eq.~\eqref{eq:ND_t>0_form_inf}, we symmetrize it with respect to integration variables $q$ and $q'$, and use the fact that $\alpha_{pq}$ and $\beta_{pq}$ are real.
Using explicit expressions of Bogoliubov coefficients \eqref{eq:alpha_beta_value_inf} and mode function \eqref{eq:g_inf}, we obtain
\begin{gather}
	\langle 0_f | {\bm T}_{\pm\pm} | 0_f \rangle_{t>0}
	=
	\frac{1}{\pi^3} \int_0^\infty dp
	\Bigg(
		\frac{1}{p}
		[ \int_0^\infty dq \cos ( qz_\pm ) + p^2 \int_0^\infty dq \frac{ \cos ( qz_\pm ) }{ q^2-p^2 }  ]^2
		+
		p [ \int_0^\infty dq \frac{ q \sin ( qz_\pm ) }{ q^2-p^2 } ]^2	 
	\Bigg).
\label{eq:ND_t>0_value1_inf}
\end{gather}
The integration over $q$ in Eq.~\eqref{eq:ND_t>0_value1_inf} can be computed to give
\be
	\langle 0_f | {\bm T}_{\pm\pm} | 0_f \rangle_{t>0}
	=
	\frac{ \delta^2 (z_\pm) }{ \pi } \int_0^\infty \frac{dp}{p}
	+
	\frac{ {\rm sgn}^2 (z_\pm) }{ 4\pi } \int_0^\infty dp p,
\label{eq:ND_t>0_value2_inf}
\ee
where ${\rm sgn}$ denotes the sign function,
\be
	{\rm sgn} (a)
	:=
	\begin{cases}
		\pm 1 & (a \gtrless 0) \\
		0 & (a=0) \\
	\end{cases}.
\ee
Note that we have used the following integration formulas,
\begin{align}
	\int_0^\infty  \cos (ax) dx
	&=\pi \delta (a),
\;\;\;
	( -\infty < a <\infty ),
\label{eq:intForm1}
\\
	\int_0^\infty \frac{ \cos (ax) }{ x^2-b^2 }dx
	&=
	-{\rm sgn} ( a ) \frac{\pi}{2b} \sin (ab),
\;\;\;
	(  -\infty < a <\infty , \; b>0),
\label{eq:intForm2}
\\
	\int_0^\infty \frac{ x \sin (ax) }{ x^2-b^2 }dx
	&=
	{\rm sgn} ( a ) \frac{\pi}{2} \cos (ab),
\;\;\;
	(  -\infty < a <\infty, \; b>0).
\label{eq:intForm3}
\end{align}
See Appendix~\ref{sec:int} for the derivation of the second and third formulas.

Let us consider the meaning of two terms in Eq.~\eqref{eq:ND_t>0_value2_inf}. The first term, the delta function squared multiplied by a divergent integral, represents the diverging flux emanating from the origin $(t,x)=(0,0)$ and  localizing on the null line $z_-=0$. The divergent factor involves the infrared divergence too since there in no  infrared cutoff introduced by finite $L$. The dependence of energy density on the delta function squared implies also the divergence of total energy emitted. 

The second term, at first glance, seems to represent an ambient divergent energy density and its vanishing on the null line emanating from the origin (note that ${\rm sgn}(0)=0$). As will be seen below, however, this is not the case. Namely, the divergence at $z_\pm \neq 0$ just represents the energy due to the zero-point oscillation just like Eq.~\eqref{eq:ND_t<0_inf}. Therefore, the regularized vacuum expectation value of energy-momentum tensor should be defined by subtracting such a diverging quantity distributing uniformly in space and time. As the result of such a subtraction, the divergence appears {\it on} the null line $z_- = 0$. Such a  renormalized vacuum expectation value of energy-momentum tensor is computed in Appendix~\ref{sec:green} with the Green-function method, which naturally involves the point-splitting regularization. The result is
\begin{align}
	\langle 0_g | {\bm T}_{\pm\pm} | 0_g  \rangle_{ t>0}^{\rm ren}
	=
	\frac{ \delta^2 (z_\pm) }{ \pi }\int_0^\infty \frac{dp}{p}
	+
	\begin{cases}
	\displaystyle \lim_{z_\pm'\to z_\pm}\frac{ 1 }{ 4\pi ( z_\pm - z_\pm' )^2 } & (z_\pm = 0) \\
	0 & (\mbox{otherwise})
	\end{cases}.
\label{VEVf_ren_t>0}
\end{align}
Here, $z_\pm$ and $z_\pm'$ are the coordinates of two points on which the Green functions are evaluated. As explained above, the second term diverges on the null line and vanishes elsewhere. Thus, there remain the two components of diverging flux even after the renormalization to propagate along the null line $z_-=0$. 

\subsubsection{From Dirichlet to Neumann}
\label{sec:DN_inf}

\begin{figure}
\begin{center}
\begin{minipage}[c]{0.8\textwidth}
\begin{center}
\includegraphics[height=5cm]{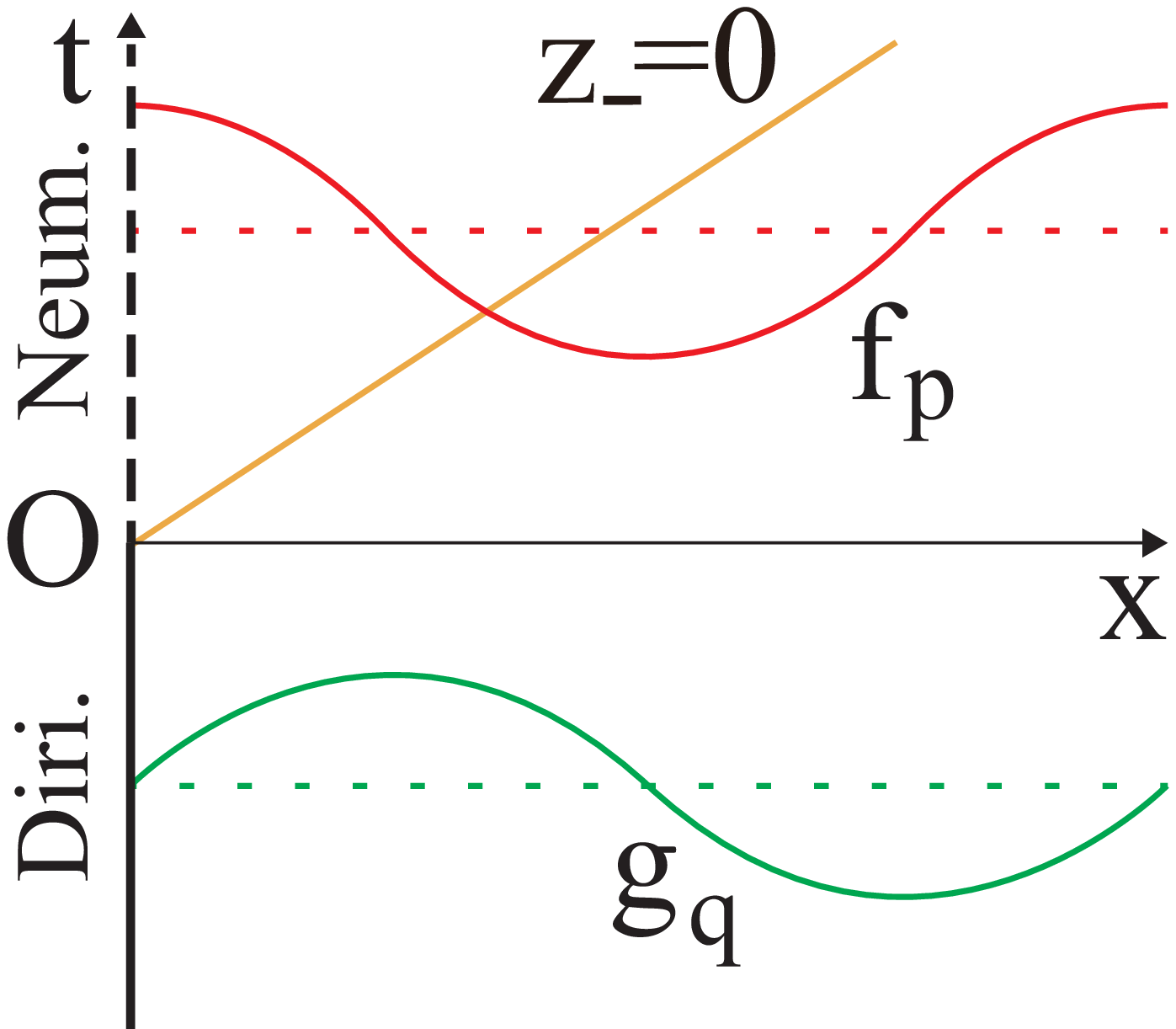}
\caption{The boundary condition at the left end of domain $(x=0)$ instantaneously changes at $t=0$ from Dirichlet (solid) to Neumann (dashed). Spatial configurations of mode functions $ f_p $ and $g_q $  are schematically depicted.}
\label{fig:DN_inf}
\end{center}
\end{minipage}
\end{center}
\end{figure}

We assume that the boundary condition at $x=0$ is Dirichlet~\eqref{eq:bc3} for $t<0$ and Neumann~\eqref{eq:bc2} for $t>0$, and that the quantum field is in vacuum $| 0_g \rangle$, given by Eq.~\eqref{eq:0g_inf}. See Fig.~\ref{fig:DN_inf} for a schematic picture of the physical situation. Then, we investigate how the vacuum is excited by computing the spectrum and energy flux of created particles.

Let us expand $g_q$ by $f_p$ as,
\be
	g_q
	=
	\int_0^\infty dp ( \rho_{qp} f_p + \sigma_{qp} f_p^\ast ).
\label{eq:gf_inf}
\ee
Here, the expansion coefficients are given by
\be
	\rho_{qp} =  \langle f_p,g_q \rangle = \alpha_{pq}^\ast,
\;\;\;
	\sigma_{qp} = - \langle f_p^\ast, g_q \rangle = - \beta_{pq},
\label{eq:rho_sigma_inf}
\ee
where $\alpha_{pq}$ and $\beta_{pq}$ are given by Eq.~\eqref{eq:alpha_beta_value_inf}.

Substituting Eq.~\eqref{eq:gf_inf} into Eq.~\eqref{eq:phi_g_inf}, and comparing it with Eq.~\eqref{eq:phi_f_inf}, we obtain
\be
	{\bm a}_p
	=
	\int_0^\infty dq ( \rho_{qp} {\bm b}_q + \sigma_{qp}^\ast {\bm b}_q^\dagger ).
\label{eq:ab_inf}
\ee
Substituting Eq.~\eqref{eq:ab_inf} into Eq.~\eqref{eq:comm_a_inf}, and using Eq.~\eqref{eq:comm_b_inf}, we obtain the unitarity relations,
\begin{align}
	\int_0^\infty dq ( \rho_{qp} \rho_{qp'}^\ast - \sigma_{qp}^\ast \sigma_{qp'} ) 
	=
	\delta ( p-p' ),
\;\;\;
	\int_0^\infty dq ( \rho_{qp} \sigma_{qp'}^\ast - \sigma_{qp}^\ast \rho_{qp'} ) 
	=
	0.
\label{eq:UR_DN_inf}
\end{align}
In Appendix \ref{sec:UR_DN_inf}, it is shown that Bogoliubov coefficients \eqref{eq:rho_sigma_inf} indeed satisfy Eq.~\eqref{eq:UR_DN_inf}.

The spectrum is computed as
\be
	\langle 0_g | {\bm a}_p^\dagger {\bm a}_p | 0_g \rangle
	=
	\int_0^\infty dq | \sigma_{qp} |^2
	=
	\frac{1}{ \pi^2 }
	\int_0^\infty dq \frac{ q }{ p (p+q)^2  },
\label{eq:aa_inf}
\ee
which is divergent. 

The expectation value of energy-momentum tensor before the change of boundary condition at $t=0$ is computed by substituting Eq.~\eqref{eq:phi_g_inf} into Eq.~\eqref{eq:em_null}, and using Eqs.~\eqref{eq:comm_b_inf} and \eqref{eq:g_inf}, as
\begin{align}
	\langle 0_g | {\bm T}_{\pm\pm} | 0_g \rangle_{t<0}
	=
	\int_0^\infty dq | \pd_\pm g_q |^2
	=
	\frac{ 1 }{ 4\pi } \int_0^\infty dq q.
\label{eq:DN_t<0_inf}
\end{align}
This represents the divergence due to the zero-point oscillation, and the regularized value vanishes as given by Eq.~\eqref{VEVg_3}.

The expectation value of energy-momentum tensor for $t>0$ is computed by substituting Eq.~\eqref{eq:phi_f_inf} into Eq.~\eqref{eq:em_null}, and using Eq.~\eqref{eq:ab_inf}, as
\begin{gather}
	\langle 0_g | {\bm T}_{\pm\pm} | 0_g \rangle_{t>0}
	=
	\int_0^\infty \int_0^\infty \int_0^\infty dq dp dp'
	[
		( \rho_{qp} \sigma_{qp'} + \rho_{qp'} \sigma_{qp} )
		{\rm Re} ( \pd_\pm f_p \pd_\pm f_{p'} )
\nn
\\
		+
		( \rho_{qp} \rho_{qp'} + \sigma_{qp} \sigma_{qp'} )
		{\rm Re} ( \pd_\pm f_p \pd_\pm f_{p'}^\ast )
	],
\label{eq:DN_t>0_form_inf}
\end{gather}
where we symmetrize it with respect to integration variables $p$ and $p'$, and use the fact that $\rho_{qp}$ and $\sigma_{qp}$ are real. Substituting explicit form of the Bogoliubov coefficients, given by Eqs.~\eqref{eq:rho_sigma_inf} and \eqref{eq:alpha_beta_value_inf}, and mode function \eqref{eq:g_inf} into Eq.~\eqref{eq:DN_t>0_form_inf}, we have
\begin{align}
	&\langle 0_g | {\bm T}_{\pm\pm} | 0_g \rangle_{t>0}
	=
	\frac{1}{\pi^3} \int_0^\infty dq
	\left(
		q^3 [ \int_0^\infty dp \frac{ \cos (pz_\pm) }{ p^2-q^2 } ]^2
		+ q [ \int_0^\infty dp \frac{ p \sin ( pz_\pm ) }{ p^2 - q^2 } ]^2
	\right).
\label{eq:DN_t>0_value1_inf}
\end{align}
The integrations over $p$ in Eq.~\eqref{eq:DN_t>0_value1_inf} are evaluated using formulas \eqref{eq:intForm2} and \eqref{eq:intForm3} to obtain
\begin{align}
	&\langle 0_g | {\bm T}_{\pm\pm} | 0_g \rangle_{t>0}
	=
	\frac{ {\rm sgn}^2 (z_\pm) }{ 4\pi } \int_0^\infty dq q.
\label{eq:DN_t>0_value2_inf}
\end{align}

Again, result \eqref{eq:DN_t>0_value2_inf} seems to represent a diverging flux and its vanishing on the null line emanating from the origin. After subtracting the uniform contribution from the zero-point oscillation, however, the divergence appears {\it on} the null line. This is explicitly shown by adopting the Green-function method in Appendix~\ref{sec:green}. The result is given by
\begin{align}
	\langle 0_g | {\bm T}_{\pm\pm} | 0_g  \rangle^{{\rm ren}}_{t>0}
	=
	\begin{cases}
	\displaystyle \lim_{z_\pm' \to z_\pm} \frac{1}{4\pi (z_\pm - z_\pm')^2 }  & (z_\pm =0) \\
	0 &  ( \mbox{otherwise} )\\
	\end{cases}.
\label{VEVg_t>0_4}
\end{align}
Here, $z_\pm$ and $z_\pm'$ are the coordinates of two points on which the Green functions are evaluated. The flux diverges on the null line and vanishes elsewhere. Thus, there remains only one component of diverging flux after the renormalization to propagate along the null line $z_-=0$. 

\section{Conclusion}
\label{sec:conc}

We have investigated the particle creation due to the instantaneous change of boundary condition (BC) in the one-dimensional (1D) finite cavity (Secs.~\ref{sec:fin} and \ref{sec:ih}) and semi-infinite cavity (Sec.~\ref{sec:inf}) by computing the vacuum expectation value of energy-momentum tensor for the free massless Klein-Gordon scalar field. The BC changes from Neumann to Dirichlet (N-D) in Secs.~\ref{sec:fin} and \ref{sec:inf}, from Neumann-Neumann to Dirichlet-Dirichlet (NN-DD) in Sec.~\ref{sec:ih}, and from Dirichlet to Neumann (D-N) in Secs.~\ref{sec:fin} and \ref{sec:inf}. 

Although any actual change of BC takes a finite interval of time, we believe that these models are capable of extracting the essence of phenomenon when the BC changes rapidly enough compared to typical time scales in the system. In particular, it is plausible that such a situation is realized for the gravitational phenomena like the appearance of strong (or wave-singular) naked singularities~\cite{Ishibashi:2002ac} and topology change of spacetime (or string) in quantum gravity~\cite{Anderson:1986ww}. In addition, the choice of Dirichlet and Neumann BCs introduced no adjustable parameters into the system, which made the whole analysis simple to be a good starting point for succeeding considerations. Most models of the particle creation due to time-dependent BCs (i.e., the dynamic Casimir effect) would have to reproduce the results in this paper in their limit of infinitely rapid change.

Thanks to the above simplifications made in our model, we could obtain almost all the results in completely analytic form. For the finite cavity N-D (resp.\ D-N) case, the vacuum expectation value of energy-momentum tensor was obtained as Eq.~\eqref{eq:ND_t>0_value6_fin} (resp.\ \eqref{eq:DN_t>0_value5_fin}). Our result that the flux in the N-D and D-N cases consist of two terms and only one term, respectively, seemed to contradict the result in Ref.~\cite{Ishibashi:2002ac}, which analyses the NN-DD case. Therefore, we revisited the NN-DD case in Sec.~\ref{sec:ih} to obtain Eq.~\eqref{eq:NNDD_t>0_value2_ih}, which is consistent with the result in Sec.~\ref{sec:fin}. The flux in the N-D and NN-DD cases consist of terms of $\delta^2(z_\pm)$ and $1/(z_\pm-z_\pm')^2$, while the flux in the D-N case consists of only term of $1/(z_\pm-z_\pm')^2 $. Although we cannot argue which term is stronger to dominate at this point, it will be the case that not only the flux but also the total energy radiated becomes large since the integration of flux cross $z_\pm=0$ diverges.

While the results in the semi-infinite cavity for the N-D case~\eqref{eq:ND_t>0_value2_inf} and D-N case~\eqref{eq:DN_t>0_value2_inf} are quite similar to their respective counterparts in the finite cavity, the analysis for the infinite cavity is much simpler than the finite-cavity case in that non-trivial mathematical formulas such as summation formulas of Eqs.~\eqref{eq:sumForm2}, \eqref{eq:maru_typo1}, and so on, are not necessary. This is a technical but an important point for succeeding studies such as the generalizations of this work (future works will be mentioned later). In addition, the vacuum expectation value of energy-momentum tensor in the semi-infinite cavity was re-derived by the Green-function method in Appendix~\ref{sec:green}. This method not only naturally involves the point-splitting regularization but also involves only simpler calculations than the Bogoliubov method in the text. Again, this is a technical but an important point. Finally, the analysis for the semi-infinite cavity confirmed that the divergence of flux due to the change of BC is nothing but an ultraviolet effect rather than an infrared one, and that the divergence of the flux has nothing to do with the Casimir effect, which exists only when $L$ is finite.

Let us discuss the origin of asymmetry between the N-D and D-N cases, of which similar conjecture was proposed in a previous paper of the present author and his collaborators~\cite{Harada:2016kkq}. The $\delta^2$-term seems to stem from a temporal discontinuity of mode function $f_n$ and $f_p$. For instance, in the finite-cavity N-D case, mode function $f_n$ is given by Eq.~\eqref{eq:f_fin} for $t<0$, having a non-zero value at $x=0$, but given by Eq.~\eqref{eq:fg_fin} for $t>0$, vanishing at $x=0$. Therefore, $f_n(t,0)$ is discontinuous as a function of time at $t=0$. On the other hand, in the finite-cavity D-N case, mode function $g_m$ is given by Eq.~\eqref{eq:g_fin} for $t<0$ and Eq.~\eqref{eq:gf_fin} for $t>0$, both of which vanish at $x=0$. Therefore, $g_m(t,0)$ is continuous as a function of time at $t=0$. In a similar way, $h_k(t,0)$ and $h_k(t,L)$ are discontinuous as functions of time at $t=0$ in the NN-DD case, and $f_p(t,0)$ (resp.\ $g_q(t,0)$) is discontinuous (resp.\ continuous) at $t=0$ in the semi-infinite N-D (resp.\ D-N) case. We conjecture that such a discontinuity, which would create a shock in the classical mechanics point of view, is the origin of the delta function squared.

Naively speaking, the results in this paper suggest that the backreaction of created particles to the spacetime and/or the cavity cannot be ignored. However, the analysis is based on the test-field approximation, therefore, it is too early to assert such an implication of the results. As a next step, it is natural to investigate the back-reaction through, say, the semi-classical Einstein equation, where the right-hand side of Einstein equation is replaced by the regularized vacuum expectation value of energy-momentum tensor of quantized fields~\cite{Birrell:1982ix}.

Given the results in this paper, there would be several directions to proceed besides investigating the back-reaction mentioned above. Firstly, it is natural to generalize the present analysis to higher-dimensional spacetime (see Ref.~\cite{Zhou:2016hsh} for a highly relevant study). Secondly, it would be important to generalize the BC in the present paper (i.e., Dirichlet and Neumann) to the Robin-type BC, which takes the form of $ \phi (t,x) - a \pd_x \phi (t,x)|_{x=0} = 0 $. Taking different values of constant $a$ before and after $t=0$, one can generalize the present analysis. By such a generalization, we would be able to verify the above conjecture about the origin of asymmetry between the N-D and D-N cases, and understand more deeply how the time-dependent BCs make the quantum vacuum excite in general.

\subsection*{Acknowledgments}
The author would like to thank T.~Harada and S.~Kinoshita for useful discussions, and anonymous referees for various suggestions to improve the early versions of manuscript. This work was partially supported by JSPS KAKENHI Grant Numbers 15K05086 and 18K03652.

\appendix
\section{Proof of unitarity relations}
\label{sec:UR}

\subsection{Equation \eqref{eq:UR_ND_fin}}
\label{sec:UR_ND_fin}

Using Eq.~\eqref{eq:alpha_beta_value_fin}, the left-hand sides of Eq.~\eqref{eq:UR_ND_fin} are written as
\begin{align}
	\sum_{\substack{ n=1 \\ n:{\rm odd}}}^\infty
	( \alpha_{nm} \alpha_{nm'}^\ast - \beta_{nm}^\ast \beta_{nm'} ) 
	&=
	\frac{ 32 (m+m') \sqrt{ mm' } }{ \pi^2 } U_{mm'},
\label{eq:UR_ND_App1_fin}
\\
	\sum_{\substack{ n=1 \\ n:{\rm odd}}}^\infty
	( \alpha_{nm} \beta_{nm'}^\ast - \beta_{nm}^\ast \alpha_{nm'} ) 
	&=
	- \frac{ 32 (m-m') \sqrt{ mm' } }{ \pi^2 } U_{mm'},
\label{eq:UR_ND_App2_fin}
\end{align}
where we define
\be
	U_{mm'}
	:=
	\sum_{\substack{ n=1 \\ n:{\rm odd}}}^\infty
	\frac{ 1 }{ [n^2-(2m)^2] [n^2-(2m')^2] }.
\label{eq:U_def_fin}
\ee
The summation over odd $n$ in Eq.~\eqref{eq:U_def_fin} can be computed to give
\be
	U_{mm'}
	=
	\frac{ \pi^2 }{ 16 ( 2m )^2 } \delta_{mm'},
\label{eq:U_value_fin}
\ee
using the following formulas~\cite[pp.\ 688--689]{maru},
\begin{align}
	\sum_{k=0}^\infty \frac{1}{(2k+1)^2-a^2}
	&=
	\frac{\pi}{4a} \tan( \frac{a \pi }{2} ),
\label{eq:sumFormUR1}
\\
	\sum_{k=0}^\infty
	\frac{1}{[(2k+1)^2-a^2]^2}
	&=
	-\frac{\pi}{8a^3}\tan (\frac{a\pi}{2}) + \frac{\pi^2}{16a^2}{\rm sec}^2( \frac{a\pi}{2} ).
\label{eq:sumFormUR2}
\end{align}
Substituting Eq.~\eqref{eq:U_value_fin} into Eqs.~\eqref{eq:UR_ND_App1_fin} and \eqref{eq:UR_ND_App2_fin}, we see Eq.~\eqref{eq:UR_ND_fin} to hold.

\subsection{Equation \eqref{eq:UR_DN_fin}}
\label{sec:UR_DN_fin}

Using Eqs.~\eqref{eq:rho_sigma_form_fin} and \eqref{eq:alpha_beta_value_fin}, the left-hand sides of Eq.~\eqref{eq:UR_DN_fin} are
\begin{align}
	\sum_{m=1}^\infty ( \rho_{mn} \rho_{mn'}^\ast - \sigma_{mn}^\ast \sigma_{mn'} ) 
	&=
	\frac{2(n+n')}{\sqrt{ nn' } \pi^2 } V_{nn'},
\label{eq:UR_DN_App1_fin}
\\
	\sum_{m=1}^\infty ( \rho_{mn} \sigma_{mn'}^\ast - \sigma_{mn}^\ast \rho_{mn'} ) 
	&=
	- \frac{2(n-n')}{\sqrt{ nn' } \pi^2 } V_{nn'},
\label{eq:UR_DN_App2_fin}
\end{align}
where we define
\begin{align}
	V_{nn'}
	&:=
	\sum_{m=1}^\infty 
	\frac{ m^2 }{ [ m^2-(n/2)^2 ] [ m^2-(n'/2)^2 ] }
\nn
\\
	&=
	\sum_{m=1}^\infty 
	\frac{ 1 }{  m^2-(n'/2)^2  }
	+
	( \frac{n}{2} )^2 
	\sum_{m=1}^\infty  \frac{ 1 }{ [ m^2-(n/2)^2 ] [ m^2-(n'/2)^2 ] }.
\label{eq:V_def_fin}
\end{align}
The summations over $m$ in Eq.~\eqref{eq:V_def_fin} can be computed to give
\be
	V_{nn'}
	=
	\frac{\pi^2}{4} \delta_{nn'},
\label{eq:V_value_fin}
\ee
using the following formulas~\cite[pp.~68--69]{iwa2},
\begin{align}
	\sum_{k=1}^\infty \frac{1}{y^2-k^2}
	&=
	\frac{\pi}{2y} \cot (\pi y) -\frac{1}{2y^2},
\label{eq:sumFormUR3}
\\
	\sum_{k=1}^\infty \frac{1}{ [ (ky)^2-1]^2}
	&=
	\frac{\pi^2}{4y^2 } {\rm cosec}^2 ( \frac{\pi}{y} ) 
	+
	\frac{\pi}{4y} \cot ( \frac{\pi}{y} ) - \frac12.
\label{eq:sumFormUR4}
\end{align}
Substituting Eq.~\eqref{eq:V_value_fin} into Eqs.~\eqref{eq:UR_DN_App1_fin} and \eqref{eq:UR_DN_App2_fin}, we see Eq.~\eqref{eq:UR_DN_fin} to hold.

\subsection{Equation~\eqref{eq:UR_ih}}
\label{sec:UR_ih}

Using Eqs.~\eqref{eq:comm_c}, \eqref{eq:xi_zeta_value1_ih}, and \eqref{eq:xi_zeta_value2_ih}, the left-hand side of Eq.~\eqref{eq:UR_ih} is written as
\begin{align}
&
	[ {\bm \xi}_m, {\bm \xi}_{m'}^\dagger ]
	+
	\sum_{k=1}^\infty ( \xi_{km} \xi_{km'}^\ast - \zeta_{km}^\ast \zeta_{km'} )
\nn
\\
&
	=
	\frac{ 8(m+m')\sqrt{ mm' } }{ \pi^2 }
	\left[
		\left( \frac{1}{2m^2m'^2} + W_{mm'} \right) \delta_{m:{\rm odd}}\delta_{m':{\rm odd}}
		+
		X_{mm'}  \delta_{m:{\rm even}}\delta_{m':{\rm even}}
	\right],
\label{eq:UR_NNDD_App1_ih}
\\
&
	[ {\bm \xi}_m, {\bm \xi}_{m'} ]
	+
	\sum_{k=1}^\infty
	\left( \xi_{km} \zeta_{km'}^\ast - \zeta_{km}^\ast \xi_{km'} \right)
\nn
\\
&	
	=
	-\frac{ 8(m-m')\sqrt{ mm' } }{ \pi^2 }
	\left[
		\left( \frac{1}{2m^2m'^2} + W_{mm'} \right) \delta_{m:{\rm odd}}\delta_{m':{\rm odd}}
		+
		X_{mm'}  \delta_{m:{\rm even}}\delta_{m':{\rm even}}
	\right],
\label{eq:UR_NNDD_App2_ih}
\end{align}
where
\begin{gather}
	W_{mm'}
	:=
	\sum_{\substack{ k=2 \\ k:{\rm even}}}^\infty
	\frac{1}{ (k^2-m^2)(k^2-m'^2) },
\;\;\;
	X_{mm'}
	:=
	\sum_{\substack{ k=1 \\ k:{\rm odd}}}^\infty
	\frac{1}{ (k^2-m^2)(k^2-m'^2) }.
\end{gather}
Applying formulas~\eqref{eq:sumFormUR3} and \eqref{eq:sumFormUR4} to $W_{mm'}$, and formulas~\eqref{eq:sumFormUR1} and \eqref{eq:sumFormUR2} to $X_{mm'}$, we obtain 
\be
	W_{mm'}
	=
	- \frac{1}{2m^2 m'^2} + \frac{\pi^2}{16m^2} \delta_{mm'} ,
\;\;\;
	X_{mm'}
	=
	\frac{\pi^2}{16m^2} \delta_{mm'}.
\label{eq:WX}
\ee
Substituting Eq.~\eqref{eq:WX} into Eqs.~\eqref{eq:UR_NNDD_App1_ih} and \eqref{eq:UR_NNDD_App2_ih}, we see that unitarity relation~\eqref{eq:UR_ih} holds.

\subsection{Equation \eqref{eq:UR_ND_inf}}
\label{sec:UR_ND_inf}

Using Eq.~\eqref{eq:alpha_beta_value_inf}, the left-hand side of Eq.~\eqref{eq:UR_ND_inf} is written as
\begin{align}
	\int_0^\infty dp ( \alpha_{pq} \alpha_{pq'}^\ast - \beta_{pq}^\ast \beta_{pq'} ) 
	&=
	\frac{ 2 (q+q') \sqrt{ qq' } }{ \pi^2 } U_{qq'},
\label{eq:UR_ND_App1_inf}
\\
	\int_0^\infty dp ( \alpha_{pq} \beta_{pq'}^\ast - \beta_{pq}^\ast \alpha_{pq'} ) 
	&=
	- \frac{ 2 (q-q') \sqrt{ qq' } }{ \pi^2 } U_{qq'},
\label{eq:UR_ND_App2_inf}
\end{align}
where we define
\be
	U_{qq'}
	:=
	\int_0^\infty  \frac{ dp }{ ( p^2-q^2 )( p^2-q^{'2} ) }.
\label{eq:U_def_inf}
\ee
By simple algebra, this is rewritten as
\be
	U_{qq'}
	=
	\frac{1}{4q(q+q')}
	\int_{-\infty}^{\infty} dp
	\left[
		\frac{1}{(p-q)(p-q')} + \frac{1}{(p+q)(p+q')} -  \frac{2}{p^2-q'^{2}}
	\right].
\label{eq:U_value1_inf}
\ee
Adapting the following formula \cite[p.\ 488]{aw} to the first and second terms of Eq.~\eqref{eq:U_value1_inf},
\be
	\int_{-\infty}^{\infty} 
	\frac{dx}{(x- a )(x- b )}  
	=
	\pi^2 \delta ( a - b ),
\;\;\;
	(-\infty < a, b < \infty),
\ee
and noting the third term vanishes from Eq.~\eqref{eq:intForm2}, we have
\be
	U_{qq'}
	=
	\frac{\pi^2}{4q^2} \delta (q-q').
\label{eq:U_value2_inf}
\ee
Substituting Eq.~\eqref{eq:U_value2_inf} into Eqs.~\eqref{eq:UR_ND_App1_inf} and \eqref{eq:UR_ND_App2_inf}, we see Eq.~\eqref{eq:UR_ND_inf} to hold.

\subsection{Equation \eqref{eq:UR_DN_inf}}
\label{sec:UR_DN_inf}

Using Eqs.~\eqref{eq:rho_sigma_inf} and \eqref{eq:alpha_beta_value_inf}, the left-hand side of Eq.~\eqref{eq:UR_DN_inf} is written as
\begin{align}
	\int_0^\infty dq ( \rho_{qp} \rho_{qp'}^\ast - \sigma_{qp}^\ast \sigma_{qp'} ) 
	&=
	\frac{2(p+p')}{ \pi^2 \sqrt{ pp' } } V_{pp'},
\label{eq:UR_DN_App1_inf}
\\
	\int_0^\infty dq ( \rho_{qp} \sigma_{qp'}^\ast - \sigma_{qp}^\ast \rho_{qp'} ) 
	&=
	- \frac{2(p-p')}{ \pi^2 \sqrt{ pp' }  } V_{pp'}.
\label{eq:UR_DN_App2_inf}
\end{align}
where we define 
\begin{align}
	V_{pp'}
	:=
	\int_0^\infty dq
	\frac{ q^2 }{ ( q^2-p^2 ) ( q^2-p^{'2} ) }.
\label{eq:V_def_inf}
\end{align}
This is computed as
\begin{align}
	V_{pp'}
	=
	\int_0^\infty dq
	\frac{ 1 }{ q^2-p^{'2} }
	+
	p^2 \int_0^\infty dq
	\frac{ 1 }{ ( q^2-p^2 ) ( q^2-p^{'2} ) }
	=
	\frac{\pi^2}{4} \delta(p-p'),
\label{eq:V_value_inf}
\end{align}
where the first term vanishes from Eq.~\eqref{eq:intForm2}, and the technique to obtain Eq.~\eqref{eq:U_value2_inf} is used to compute the second term. Substituting Eq.~\eqref{eq:V_value_inf} into Eqs.~\eqref{eq:UR_DN_App1_inf} and \eqref{eq:UR_DN_App2_inf}, we see Eq.~\eqref{eq:UR_DN_inf} to hold.

\section{Green-function method for semi-infinite cavity}
\label{sec:green}

We re-analyze the vacuum excitation by the change of boundary condition for the semi-infinite cavity using the Green-function method~\cite{Birrell:1982ix, Harada:2016kkq}, which naturally incorporates the renormalization of zero-point energy. 

\subsection{Green functions}

Two Hadamard elementary functions, $F^{(1)}$ and $G^{(1)}$, are defined by
\begin{align}
	F^{(1)}(z, z')
	:=
	\langle 0_f | \{  {\bm \phi}( z), {\bm \phi}( z' ) \} | 0_f \rangle,
\;\;\;
	G^{(1)}(z, z')
	:=
	\langle 0_g | \{  {\bm \phi}( z), {\bm \phi}( z' ) \} | 0_g \rangle,
\label{F1G1}
\end{align}
where we have introduced a simplified notation $z:=(z_-,z_+)$ and $z':=(z_-',z_+')$, and $\{ \cdot, \cdot \}$ denotes the anti-commutator, $\{ \phi, \psi \} := \phi \psi + \psi \phi$. Two Pauli-Jordan or Schwinger functions, $F$ and $G$, are defined by
\begin{align}
	i F (z, z')
	:=
	\langle 0_f | [  {\bm \phi}( z), {\bm \phi}( z' ) ] | 0_f \rangle,
\;\;\;
	i G (z, z')
	:=
	\langle 0_g | [  {\bm \phi}( z), {\bm \phi}( z' ) ] | 0_g \rangle.
\label{iFiG}
\end{align}

Using the decompositions of field operator~\eqref{eq:phi_f_inf} and \eqref{eq:phi_g_inf}, the Hadamard elementary functions are represented as
\begin{align}
	F^{(1)}(z, z')
	&=
	\int_0^\infty dp [ f_p(z) f_p^\ast (z') + {\rm c.c.} ]
\label{F1a}
\\
	&=
	\frac{1}{2\pi}
	\int_0^\infty \frac{dp}{p}
	[
		\cos (p\Delta z_-) + \cos (p\Delta z_+) + \cos p(z_- - z_+') + \cos p (z_+- z_-')
	],
\label{F1b}
\\
	G^{(1)}(z, z')
	&=
	\int_0^\infty dq [ g_q(z) g_q^\ast (z') + {\rm c.c.} ]
\label{G1a}
\\
	&=
	\frac{1}{2\pi}
	\int_0^\infty \frac{dq}{q}
	[
		\cos (q\Delta z_-) + \cos (q\Delta z_+) - \cos q(z_- - z_+') - \cos q (z_+- z_-')
	],
\label{G1b}
\end{align}
where ${\rm c.c.}$ denotes the complex conjugate and $ \Delta z_\pm := z_\pm - z_\pm' $. For the Pauli-Jordan functions, the momentum integration can be evaluated to give
\begin{align}
	iF(z,z')
	&=
	-\frac{i}{4}
	[ {\rm sgn} (\Delta z_-) + {\rm sgn} (\Delta z_+) + {\rm sgn} ( z_- - z_+') + {\rm sgn} ( z_+ - z_-') ],
\label{iF}
\\
	iG(z, z')
	&=
	-\frac{i}{4}
	[ {\rm sgn} (\Delta z_-) + {\rm sgn} (\Delta z_+) - {\rm sgn} ( z_- - z_+') - {\rm sgn} ( z_+ - z_-') ],
\label{iG}
\end{align}
where we have used $ 	\int_0^\infty \frac{\sin (ax)}{x} dx = \pm \frac{\pi}{2} \; (a \gtrless 0)$~\cite[p.~251]{iwa1}.

\subsection{From Neumann to Dirichlet}

The vacuum expectation value of energy-momentum tensor before the change of boundary condition is obtained by differentiating the Hadamard elementary function $F^{(1)}$ with respect to two points $z$ and $z'$, and taking the same-point limit $z' \to z$,
\begin{align}
	\langle 0_f | {\bm T}_{\pm\pm} | 0_f  \rangle_{t<0}^{\rm Green}
	=
	\frac12 \lim_{z' \to z} \pd_\pm \pd_{\pm}' F^{(1)}(z,z').
\label{VEVf_1}
\end{align}
From Eqs.~\eqref{F1b} and \eqref{VEVf_1}, one obtains
\begin{align}
	\langle 0_f | {\bm T}_{\pm\pm} | 0_f  \rangle_{t<0}^{\rm Green}
	&=
	 \lim_{z' \to z} \frac{1}{4\pi} \int_0^\infty dp p \cos (p\Delta z_\pm)
\label{VEVf_2_pre}
\\
	&=
	\lim_{z' \to z} \frac{-1}{ 4\pi ( \Delta z_\pm )^2 }.
\label{VEVf_2}
\end{align}
One can see that Eq.~\eqref{VEVf_2_pre} reproduces Eq.~\eqref{eq:ND_t<0_inf} if one takes limit $z' \to z$ before the $p$-integration. Equation~\eqref{VEVf_2} shows that $ \langle 0_f | {\bm T}_{\pm\pm} | 0_f  \rangle_{t<0}^{\rm Green} $ contains the ultraviolet divergence $\sim 1/( \Delta z_\pm )^2$, which is the vacuum energy due to the zero-point oscillation always existing even in a free Mankowski spacetime. Therefore, the renormalized energy-momentum is defined by subtracting this ultraviolet divergence as
\begin{align}
	\langle 0_f | {\bm T}_{\pm\pm} | 0_f  \rangle_{ t<0}^{\rm ren}
	&:=
	\langle 0_f | {\bm T}_{\pm\pm} | 0_f  \rangle_{t<0}^{\rm Green}
	-
	\lim_{z' \to z} \frac{-1}{ 4\pi ( \Delta z_\pm )^2 }
\label{VEVf_ren}
\\
	&=
	0,
\label{VEVf_ren2}
\end{align}
which reasonably vanishes before changing the boundary condition.

The vacuum expectation value of energy-momentum tensor after the change the  boundary condition has the same expression as Eq.~\eqref{VEVf_1}. However, since the boundary condition is changed at $t=0$, Hadamard elementary function $F^{(1)}$ before the change of boundary condition has to be propagated into $t>0$ region using Pauli-Jordan function $iG$~\cite{Harada:2016kkq}. Thus, the energy-momentum is represented as
\begin{align}
	\langle 0_f | {\bm T}_{\pm\pm} | 0_f  \rangle_{t>0}^{\rm Green}
	=
	\frac12 \lim_{B \to A} \pd_\pm \pd_{\pm}' [ iG(A,C)  iG(B, D)  F^{(1)} (C, D) ],
\label{VEVf_t>0_1}
\end{align}
where $A:=z$ and $B:=z'$. Namely, in this abbreviated notation, let a capital Latin letters (except $G$ and $F$) denote a world point, e.g., $\phi(A, B)=\phi(z,z')$. In addition, let a pair of repeated capital Latin letter denote the Klein-Gordon inner product at $t=0$, e.g., $\phi(A) \psi(A) := \langle \phi, \psi \rangle|_{t=0} $. 

Substituting Eq.~\eqref{F1a} into Eq.~\eqref{VEVf_t>0_1}, one obtains
\begin{align}
	\langle 0_f | {\bm T}_{\pm\pm} | 0_f  \rangle_{t>0}^{\rm Green}
	=
	\frac12 \lim_{B \to A} \int_0^\infty dp
	\Big(
	i \pd_\pm G(A,C)f_p(C)
	[  i \pd_\pm' G(B,D)f_p(D) ]^\ast
	+ {\rm c.c.}
	\Big),
\label{VEVf_t>0_2}
\end{align}
where we have used the property of inner product $\langle \phi, \psi^\ast \rangle = - \langle \phi^\ast , \psi \rangle^\ast$. The inner product in Eq.~\eqref{VEVf_t>0_2} can be written as
\begin{align}
	i \pd_\pm G(A,B)f_p(B)
	=\int_0^\infty dx'
	[
		\pd_\pm G(z,z') \pd_{t'} f_p(z') - \pd_{t'} \pd_\pm G(z,z') f_p(z')
	]|_{t'=0}.
\label{Gf1}
\end{align}
Using Eq.~\eqref{iG}, derivatives of $G$ in Eq.~\eqref{Gf1} are computed as
\begin{align}
	\pd_\pm G(z,z')|_{t'=0}
	&=
	-\frac12 [ \delta(x' \mp z_{\pm}) - \delta( x'\pm z_\pm )],
\label{Gp}
\\
	\pd_{t'} \pd_\pm G(z,z')|_{t'=0}
	&=
	\mp \frac12 \pd_{x'} [ \delta(x' \mp z_{\pm}) + \delta( x'\pm z_\pm )],
\label{Gpp}
\end{align}
where $ {\rm sgn}'(x)=   2\delta(x) $ was used. Substituting Eqs.~\eqref{Gp} and \eqref{Gpp} into Eq.~\eqref{Gf1}, one obtains
\begin{align}
	i \pd_\pm G(A,B)f_p(B)
	=
	\pm \frac{i}{2} \sqrt{\frac{p}{\pi}} {\rm sgn}(z_\pm) e^{-ipz_{\pm}}
	\mp \frac{\delta(z_\pm)}{\sqrt{\pi p}},
\label{Gf2}
\end{align}
where we have used 
\begin{align}
	\int_{0}^\infty  \delta (x-a) f(x) dx &= \theta(a) f(a),
\label{delta-form}
\\
	\theta( \pm x ) - \theta (\mp x) &= \pm {\rm sgn} (x).
\label{theta-form}
\end{align}
With Eq.~\eqref{Gf2}, Eq.~\eqref{VEVf_t>0_2} yields
\begin{align}
	\langle 0_f | {\bm T}_{\pm\pm} | 0_f  \rangle_{t>0}^{\rm Green}
	=
	\frac{ \delta^2 (z_\pm) }{ \pi }\int_0^\infty \frac{dp}{p}
	+
	{\rm sgn}^2(z_\pm) \lim_{z'\to z}\frac{ -1 }{ 4\pi ( \Delta z_\pm )^2 }.
\label{VEVf_t>0_3}
\end{align}
The last term in Eq.~\eqref{VEVf_t>0_3} shows that $  \langle 0_g | {\bm T}_{\pm\pm} | 0_g  \rangle_{t>0}^{\rm Green} $ contains the ultraviolet divergence due to zero-point oscillation. Thus, the renormalized energy-momentum is defined in the same way as Eq.~\eqref{VEVf_ren} by subtracting the zero-point energy,
\begin{align}
	\langle 0_f | {\bm T}_{\pm\pm} | 0_f  \rangle_{ t>0}^{\rm ren}
	&:=
	\langle 0_f | {\bm T}_{\pm\pm} | 0_f  \rangle_{t>0}^{\rm Green}
	-
	\lim_{z'\to z}\frac{ -1 }{ 4\pi ( \Delta z_\pm )^2 }
\\
	&=
	\frac{ \delta^2 (z_\pm) }{ \pi }\int_0^\infty \frac{dp}{p}
	+
	\begin{cases}
	\displaystyle \lim_{z'\to z}\frac{ 1 }{ 4\pi ( \Delta z_\pm )^2 } & (z_\pm = 0) \\
	0 & (\mbox{otherwise})
	\end{cases},
\label{VEVf_ren_t>0_pre}
\end{align}
which is nothing but Eq.~\eqref{VEVf_ren_t>0}.

\subsection{From Dirichlet to Neumann}

The vacuum expectation value of energy-momentum tensor before the change of boundary condition is given by 
\begin{align}
	\langle 0_g | {\bm T}_{\pm\pm} | 0_g  \rangle_{t<0}^{\rm Green}
	=
	\frac12 \lim_{z' \to z} \pd_\pm \pd_{\pm}' G^{(1)}(z,z').
\label{VEVg_1}
\end{align}
Substituting Eq.~\eqref{G1b} into Eq.~\eqref{VEVg_1}, one obtains 
\begin{align}
	\langle 0_g | {\bm T}_{\pm\pm} | 0_g  \rangle_{t<0}^{\rm Green}
	=
	\lim_{z' \to z} \frac{1}{4\pi} \int_0^\infty dq q \cos ( q\Delta z_\pm )
	=
	\lim_{z' \to z} \frac{-1}{ 4\pi ( \Delta z_\pm )^2 }.
\label{VEVg_2}
\end{align}
This represents the ultraviolet divergence due to the zero-point oscillation. The normalized energy-momentum is defined by subtracting such a divergence,
\begin{align}
	\langle 0_g | {\bm T}_{\pm\pm} | 0_g  \rangle_{ t<0}^{\rm ren}
	:=
	\langle 0_g | {\bm T}_{\pm\pm} | 0_g  \rangle_{t<0}^{\rm Green}
	-
	\lim_{z' \to z} \frac{-1}{ 4\pi ( \Delta z_\pm )^2 }
	=
	0,
\label{VEVg_3}	
\end{align}
which reasonably vanishes before the change of boundary condition.

The energy-momentum after the change of boundary condition is obtained by propagating $G^{(1)}$ by $iF$,
\begin{align}
	\langle 0_g | {\bm T}_{\pm\pm} | 0_g  \rangle_{t>0}^{\rm Green}
	=
	\frac12 \lim_{B \to A} \pd_\pm \pd_{\pm}' [ iF(A,C)  iF(B, D)  G^{(1)} (C, D) ].
\label{VEVg_t>0_1}
\end{align}
Substituting Eq.~\eqref{G1a}, this quantity is represented as
\begin{align}
	\langle 0_g | {\bm T}_{\pm\pm} | 0_g  \rangle_{t>0}^{\rm Green}
	=
	\frac12 \lim_{B \to A} \int_0^\infty dq
	\Big(
	i \pd_\pm F(A,C)g_q(C)
	[  i \pd_\pm' F(B,D)g_q(D) ]^\ast
	+ {\rm c.c.}
	\Big).
\label{VEVg_t>0_2}
\end{align}
The inner product in Eq.~\eqref{VEVg_t>0_2} is written as
\begin{align}
	i \pd_\pm F(A,B)g_q(B)
	&=\int_0^\infty dx'
	[
		\pd_\pm F(z,z') \pd_{t'} g_q(z') - \pd_{t'} \pd_\pm F(z,z') g_q(z')
	]|_{t'=0}.
\label{Fg1}
\end{align}
Using Eq.~\eqref{iF}, derivatives of $F$ in Eq.~\eqref{Fg1} are computed as
\begin{align}
	\pd_\pm F(z,z')|_{t'=0}
	&=
	-\frac12 [ \delta(x' \mp z_{\pm}) + \delta( x'\pm z_\pm )],
\label{Fp}
\\
	\pd_{t'} \pd_\pm F(z,z')|_{t'=0}
	&=
	\mp \frac12 \pd_{x'} [ \delta(x' \mp z_{\pm}) - \delta( x'\pm z_\pm )].
\label{Fpp}
\end{align}
Substitution of Eqs.~\eqref{Fp} and \eqref{Fpp} into Eq.~\eqref{Fg1} yields
\begin{align}
	i \pd_\pm F(A,B)g_q(B)
	=
	-\frac12 \sqrt{ \frac{q}{\pi} } {\rm sgn}(z_\pm) e^{ -iqz_{\pm} },
\label{Fg2}
\end{align}
where we have used formulas~\eqref{delta-form} and \eqref{theta-form}. The combination of Eqs.~\eqref{Fg2} and \eqref{VEVg_t>0_2} gives
\begin{align}
	\langle 0_g | {\bm T}_{\pm\pm} | 0_g  \rangle_{t>0}^{\rm Green}
	=
	{\rm sgn}^2(z_\pm) \lim_{z'\to z}\frac{ -1 }{ 4\pi ( \Delta z_\pm )^2 }.
\label{VEVg_t>0_3}
\end{align}

The renormalized energy-momentum is obtained by subtracting the zero-point energy \eqref{VEVg_2} from Eq.~\eqref{VEVg_t>0_3},
\begin{align}
	\langle 0_g | {\bm T}_{\pm\pm} | 0_g  \rangle^{{\rm ren}}_{t>0}
	&:=
	\langle 0_g | {\bm T}_{\pm\pm} | 0_g  \rangle_{t>0}^{\rm Green} - \lim_{z' \to z} \frac{-1}{4\pi (\Delta z_\pm)^2 } 
\nn
\\
	&=
	\begin{cases}
	\displaystyle \lim_{z' \to z} \frac{1}{4\pi (\Delta z_\pm)^2 }  & (z_\pm =0) \\
	0 &  ( \mbox{otherwise} )\\
	\end{cases},
\label{VEVg_t>0_4_pre}
\end{align}
which is nothing but Eq.~\eqref{VEVg_t>0_4}.

\section{Integral formulas~\eqref{eq:intForm2} and \eqref{eq:intForm3}}
\label{sec:int}

\begin{figure}
\begin{center}
\begin{minipage}[c]{0.8\textwidth}
\begin{center}
\includegraphics[height=4cm]{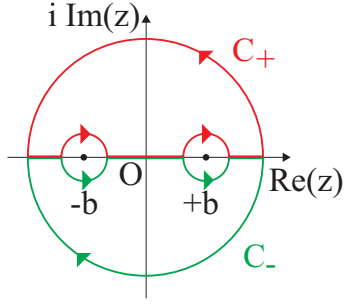}
\caption{Two closed contours $C_+$ and $C_-$ in the complex plane, each of which contains an infinitely large semicircle and two infinitesimal semicircles to avoid $-b$ and $+b$ on the real axis.}
\label{fig:cont}
\end{center}
\end{minipage}
\end{center}
\end{figure}

Let us calculate the principal values of following integrals,
\begin{align}
	I
	:=
	\int_0^\infty \frac{ \cos (ax) }{ x^2-b^2 }dx,
\;\;\;
	J
	:=
	\int_0^\infty \frac{ x \sin (ax) }{ x^2-b^2 }dx,
\label{eq:IJ_def}
\end{align}
where $ -\infty < a < \infty, \; b>0 $. Note that we always consider only principal values for improper integrals. These are written as
\begin{align}
	I
	&=
	\frac14
	( I_+ + I_- ),
\;\;\;
	I_\pm
	:=
	\int_{-\infty}^{\infty} \frac{ e^{\pm i a x} }{ x^2-b^2 } dx,
\label{eq:I_dec}
\\
	J
	&=
	\frac{1}{4i}( J_+ - J_- ),
\;\;\;
	J_\pm
	:=
	\int_{-\infty}^\infty \frac{ xe^{ \pm i a x } }{ x^2-b^2 }.
\label{eq:J_dec}
\end{align}

We suppose two contours $C_+$ and $C_-$ drawn in Fig.~\ref{fig:cont} and use Cauchy's integral theorem and the residue theorem. 

For $a>0$, taking contour $C_\pm $ for $I_\pm$ and $J_\pm$, we have
\begin{align}
	0
	&=
	\int_{C_\pm} 
	\frac{ e^{\pm i a z} }{ z^2-b^2 }dz
	=
	I_\pm  \mp \frac12 \cdot 2\pi i {\rm Res}[ I_\pm ,-b ]  \mp \frac12 \cdot 2\pi i {\rm Res}[ I_\pm, b],
\label{eq:I_Cauchy_a>0}
\\
	0
	&=
	\int_{C_\pm} \frac{ z e^{ \pm i a z } }{ z^2-b^2 } dz
	=
	J_\pm \mp \frac12 \cdot 2\pi i {\rm Res}[ J_\pm, -b ] \mp \frac12 \cdot 2\pi i {\rm Res}[ J_\pm, b ].
\label{eq:J_Cauchy_a>0}
\end{align}
Here, ${\rm Res}[X,z_0]$ denotes the residue of integrand of $X$ at $z=z_0$, and the contributions from the large semicircles vanish from Jordan's lemma. Substituting the following values of residues,
\begin{gather}
	{\rm Res}[ I_\pm , -b ] = - \frac{ e^{ \mp iab } }{ 2b },
\;\;\;
	{\rm Res}[ I_\pm , b ] =  \frac{ e^{ \pm iab } }{ 2b },
\;\;\;
{\rm Res}[ J_\pm , -b ] = \frac{e^{ \mp i ab }}{2} ,
\;\;\;
	{\rm Res}[ J_\pm , b ] = \frac{e^{ \pm i ab }}{2}
\label{eq:IJ_res}
\end{gather}
into Eqs.~\eqref{eq:I_Cauchy_a>0} and \eqref{eq:J_Cauchy_a>0}, we have
\be
	I_\pm = -\frac{\pi}{b} \sin (ab),
\;\;\;
	J_\pm = \pm i \pi \cos (ab),
\;\;\;
	(a>0).
\label{eq:IJ_value_a>0}
\ee

For $a<0$, taking contour $C_\mp$ for $I_\pm$ and $J_\pm$, we have
\begin{align}
	0
	&=
	\int_{C_\mp} 
	\frac{ e^{\pm i a z} }{ z^2-b^2 }dz
	=
	I_\pm \pm \frac12 \cdot 2\pi i {\rm Res}[ I_\pm ,-b ]  \pm \frac12 \cdot 2\pi i {\rm Res}[ I_\pm, b],
\label{eq:I_Cauchy_a<0}
\\
	0
	&=
	\int_{C_\mp} \frac{ z e^{ \pm i a z } }{ z^2-b^2 } dz
	=
	J_\pm \pm \frac12 \cdot 2\pi i {\rm Res}[ J_\pm, -b ] \pm \frac12 \cdot 2\pi i {\rm Res}[ J_\pm, b ].
\label{eq:J_Cauchy_a<0}
\end{align}
Using Eq.~\eqref{eq:IJ_res} again, we have
\be
	I_\pm = \frac{\pi}{b} \sin (ab),
\;\;\;
	J_\pm = \mp i \pi \cos (ab),
\;\;\;
	(a<0).
\label{eq:IJ_value_a<0}
\ee

For $a=0$, taking either $C_+$ or $C_-$ for $I_\pm$, one can show that $I_\pm$ vanishes. In addition,  $J$ obviously vanishes by definition~\eqref{eq:IJ_def}. Thus, we have 
\be
	I_\pm = J = 0,
\;\;\;
	(a=0).
\label{eq:IJ_value_a=0}
\ee

Combining Eqs.~\eqref{eq:I_dec}, \eqref{eq:J_dec}, \eqref{eq:IJ_value_a>0}, \eqref{eq:IJ_value_a<0}, and \eqref{eq:IJ_value_a=0}, we see formulas~\eqref{eq:intForm2} and \eqref{eq:intForm3} to hold.


\end{document}